\documentclass{aa}  

\usepackage[switch]{lineno}
\usepackage{graphicx}
\usepackage{natbib}
\usepackage{txfonts}
\usepackage[colorlinks=true,
            citecolor=blue,
            linkcolor=blue,
            urlcolor=blue]{hyperref}
\begin{document}

\title{Radiation Pressure Instability as a Possible Driver of Variability in the ``turn-on'' Changing-Look AGN SDSS J1430+2303}

\author{Han He\inst{1}
        \and Bei You\inst{1}\thanks{Corresponding author}
        \and Marzena {\'S}niegowska\inst{2}
        \and Bo{\.z}ena Czerny\inst{3}
        \and Claudio Ricci\inst{4,5,6}
        \and Xinwu Cao\inst{7}
       }

\institute{
Department of Astronomy, School of Physics,
Wuhan University, Wuhan 430072, People's Republic of China
\and
Astronomical Institute, Czech Academy of Sciences,
Bo\v{c}n{\'i} II 1401, Prague, 14100, Czechia
\and
Center for Theoretical Physics, Polish Academy of Sciences,
Al. Lotnik{\'o}w 32/46, PL-02-668 Warsaw, Poland
\and
Department of Astronomy, University of Geneva, 
ch. d’Ecogia 16, 1290 Versoix, Switzerland
\and
Instituto de Estudios Astrof{\'i}sicos, Facultad de Ingenier{\'i}a y Ciencias, 
Universidad Diego Portales, Av. Ej{\'e}rcito Libertador 441, Santiago, Chile
\and
Kavli Institute for Astronomy and Astrophysics, 
Peking University, Beijing 100871, People's Republic of China
\and
Institute for Astronomy, School of Physics, 
Zhejiang University, Hangzhou 310058, People’s Republic of China
\email{youbei@whu.edu.cn}
}

 
\abstract
{}
{We aim to investigate the multi-wavelength variability, spectral, and timing properties of the changing-look active galactic nucleus (CL AGN) SDSS J1430+2303, and to explore the possible physical origin of its peculiar variability pattern.}
{We perform a multi-wavelength analysis using optical, ultraviolet, and X-ray observations. 
We investigate the long-term optical color variation, characterize the evolution of the X-ray spectral and timing properties, and explore the possible origin of the observed optical oscillations using a radiation pressure instability model.}
{The optical flux increased by about 1 mag {within at most 5 years}, accompanied by a spectral transition from Seyfert 1.9 to 1.2. 
The long-term color variation follows the ``bluer-when-brighter'' trend, with a color-magnitude slope consistent with previous statistical results for CL and Type 1 AGNs. 
During the brightened high state, the optical light curve exhibits rapidly decaying periods with progressively decreasing amplitudes, a behavior not previously reported in other CL AGNs.
The UV and X-ray emission show more dramatic variability, with UV flares reaching $\sim0.5$ mag and the X-ray flux varying by up to an order of magnitude. 
X-ray spectral analysis reveals a weak soft excess {for most observations}, with its strength and the photon index positively correlated with the Eddington ratio, broadly following the relations reported for Type 1 AGNs. 
X-ray timing analysis shows a nearly constant hard lag at $\sim10^{-4}$ Hz during the luminosity decline, suggesting a stable disk-corona geometry. 
{Finally, a toy model with a prescribed shrinking unstable zone qualitatively reproduces the decreasing durations and amplitudes of the optical flares, suggesting radiation pressure instability as a possible, though not yet established, explanation for the observed oscillations.}}
{}


   \maketitle
%

\section{Introduction}

{Active galactic nuclei (AGN), powered by accretion onto supermassive black holes (SMBHs), are characterized by strong variability across the broadband spectrum from the infrared to X-rays (e.g., IR: \citealt{Edelson1987ApJ...323..516E,Kozlowski2016ApJ...817..119K,Sanchez2017ApJ...849..110S,Lyu2019ApJ...886...33L}; optical: \citealt{Ulrich1997ARA&A..35..445U,Kawaguchi1998ApJ...504..671K,Sesar2007AJ....134.2236S,Stone2022MNRAS.514..164S}; X-ray: \citealt{Ulrich1997ARA&A..35..445U,Lawrence1993ApJ...414L..85L,Gonzalez2012A&A...544A..80G,Vagnetti2016A&A...593A..55V,Paolillo2017MNRAS.471.4398P}). AGN vary on timescales from hours to decades, with amplitudes that depend on wavelength, luminosity, and accretion rate \citep{Ulrich1997ARA&A..35..445U,MacLeod2010ApJ...721.1014M,Kelly2009ApJ...698..895K}. This variability provides a direct probe of the structure and physical processes of the inner accretion flow, which are otherwise spatially unresolved \citep{Uttley2002MNRAS.332..231U,Vaughan2003MNRAS.345.1271V,McHardy2006Natur.444..730M,Cackett2021iSci...24j2557C}. In the standard picture, the optically thick, geometrically thin accretion disk produces the optical/UV continuum \citep{Shakura1973A&A....24..337S}. The X-ray emission arises in a hot corona, and the soft excess is attributed to warm Comptonization or to relativistic reflection \citep{Fabian2009Natur.459..540F,Done2012MNRAS.420.1848D}. Broad-band spectral models combining these components are now widely used to describe the accretion flow \citep{Kubota2018MNRAS.480.1247K}. Stochastic fluctuations in the accretion flow, reprocessing of X-rays by the disk, and propagating fluctuations are commonly invoked to explain the ubiquitous low-amplitude variability \citep{Uttley2002MNRAS.332..231U,Kelly2009ApJ...698..895K,MacLeod2010ApJ...721.1014M,Edelson2015ApJ...806..129E}. However, these processes cannot readily account for the large, coherent changes in luminosity seen in a growing number of sources \citep{Lawrence2018NatAs...2..102L}.}

{Changing-look AGN (CL AGN) represent some of the most extreme manifestations of this large-amplitude variability. Over months to years, they show dramatic changes in continuum luminosity, accompanied by the appearance or disappearance of broad emission lines, leading to a transition between Type 1 and Type 2 spectral classes \citep{Tohline1976ApJ...210L.117T,Denney2014ApJ...796..134D,Shappee2014ApJ...788...48S,LaMassa2015ApJ...800..144L,MacLeod2016MNRAS.457..389M}. Large samples assembled from multi-epoch spectroscopic surveys now comprise hundreds of objects \citep{Ruan2016ApJ...826..188R,Sheng2017ApJ...846L...7S,Yang2018ApJ...862..109Y,Stern2018ApJ...864...27S,MacLeod2019ApJ...874....8M,Frederick2019ApJ...883...31F,Graham2020PhRvL.124y1102G,Ricci2023NatAs...7.1282R,Komossa2026AdSpR..77.4041K}. The physical origin of CL AGN variability is generally discussed in terms of two broad scenarios: changes in line-of-sight obscuration and intrinsic changes in the accretion flow. In the former scenario, variations in the intervening absorbing material can produce apparent changes in the observed continuum and emission-line properties \citep{Matt2003MNRAS.342..422M,Elitzur2009ApJ...701L..91E,Elitzur2014MNRAS.438.3340E}. In contrast, intrinsic variability involves changes in the accretion rate, accretion state, or disk-corona structure, which can modify the ionizing continuum and consequently the broad-line emission \citep{Narayan1994ApJ...428L..13N,Esin1997ApJ...489..865E,Ho2008ARA&A..46..475H,Noda2018MNRAS.480.3898N,Ricci2020ApJ...898L...1R}. Among these possibilities, radiation pressure instability has attracted particular attention because it can produce large-amplitude limit-cycle variability in the inner accretion disk \citep{Pringle1976MNRAS.177...65P,Sniegowska2020A&A...641A.167S,Sniegowska2023A&A...672A..19S}. However, the applicability of this mechanism to individual CL AGN and its ability to reproduce their diverse variability patterns remain uncertain.}

{SDSS J1430 (SDSS J143016.05+230344.4; also known as tick-tock or AT2019cuk) is a nearby Seyfert 1.9 galaxy at $z=0.08105$ \citep{Stern2012MNRAS.426.2703S,Oh2015ApJS..219....1O,Jiang2022arXiv220111633J}. After many years in a low-luminosity AGN state with a low Eddington ratio ($\sim10^{-4}$) \citep{Jiang2022arXiv220111633J}, the source brightened by more than one magnitude in 2018 and developed strong broad Balmer lines, becoming a CL AGN \citep{Jiang2022arXiv220111633J,Hoshi2024PASJ...76..103H}. Its subsequent optical light curve showed quasi-periodic oscillations whose period shortened from about one year to three months while the amplitude decreased, a ``chirping'' behavior unprecedented in AGN \citep{Jiang2022arXiv220111633J}. This pattern was first ascribed to an eccentric SMBHB in which the secondary repeatedly crosses the primary disk \citep{Jiang2022arXiv220111633J,Zhong2023SCPMA..6630411Z,Dou2022A&A...665L...3D}, with coalescence expected within about three years \citep{Jiang2022arXiv220111633J}. Motivated by this prediction and the unusual variability of the source, an extensive multi-wavelength monitoring campaign was subsequently initiated, including XMM-Newton \citep{Jansen2001A&A...365L...1J}, Neil Gehrels Swift Observatory (Swift; \citealt{Gehrels2004ApJ...611.1005G}), Neutron star Interior Composition Explorer (NICER; \citealt{Gendreau2016SPIE.9905E..1HG}), Very Long Baseline Array (VLBA; \citealt{Napier1994IEEEP..82..658N}), European Very Long Baseline Interferometry (EVN; \citealt{Szomoru2004evn..conf..257S}), and the Lijiang 2.4-m telescope \citep{Wang2019RAA....19..149W}. These observations aimed to track the evolution of the source across different wavelengths and investigate the physical origin of its unusual variability. Subsequent observations disfavor this scenario. The oscillation amplitude weakened in a way inconsistent with a simple disk-crossing model \citep{Dotti2023MNRAS.518.4172D}, the optical polarization is consistent with a single-SMBH Seyfert 1 \citep{Marin2023A&A...673A.126M}, and the predicted merger has not occurred. Meanwhile, multi-wavelength follow-up revealed a compact radio core \citep{McMahon2002ApJS..143....1M,An2022A&A...663A.139A} and recurrent hard X-ray flares with very hard spectra ($\Gamma\sim1$) \citep{Masterson2023ApJ...945L..34M}. The physical origin of the variability therefore remains unexplained, and no model has yet accounted for both the shortening of the timescale and the damping of the amplitude. }

Therefore, in this paper, we present a comprehensive multi-wavelength spectral and variability analysis of SDSS J1430, combining archival X-ray, UV, and optical observations, to investigate the spectral and temporal properties of SDSS J1430, to constrain its accretion state, and to explore the physical origin of its unusual variability. The manuscript is organized as follows. In Section \ref{sec:data}, we describe the observations and data reduction. Section \ref{sec:results} presents the multi-wavelength light curves, color-magnitude diagram, X-ray spectra and timing analysis. In Section \ref{sec:discussion}, we discuss the spectral type transition, the relation between the photon index, the soft excess and the Eddington ratio. We also discuss the possible physical interpretation for the variability. We summarize our conclusions in Section \ref{sec:conclusions}. Throughout this work, we adopt a flat $\Lambda$CDM cosmology with $H_0=70~\rm km~s^{-1}~Mpc^{-1}$, $\Omega_{\rm m}=0.3$, and $\Omega_{\Lambda}=0.7$.

\section{Observations and data reduction}

\label{sec:data}

\subsection{XMM-Newton}
XMM-Newton \citep{Jansen2001A&A...365L...1J} performed seven Director's Discretionary Time (DDT) observations of SDSS J1430 in imaging mode from 2021-12-31 to 2022-08-10. We reduced the XMM-Newton EPIC data using the Python interface to the XMM-Newton Science Analysis System (pysas, SAS v21.0.0, \citealt{Valencic2025AAS...24547107V}). Only data from the PN instrument were considered because it has higher sensitivity than the MOS cameras. All the observations were carried out with the thin filter in Prime Large Window mode, except the first two observations in Prime Full Window mode. We followed the standard data reduction procedures to create the calibrated event files using {\tt epproc} with the latest calibrations. After removing the bad pixels, we screened for the high-flaring background time intervals exceeding 0.6 cts/s with single events (PATTERN==0) in the 10--12 keV band. The resultant net exposure times for each observation are listed in Table \ref{tab:pn_obsids}. Only the single events are used for the analysis. We then extract the source spectra and light curves within a circular region with a radius of 35$^{\prime\prime}$, and background products from an off-source region on the same CCD chip with the same radius. The response files are produced using {\tt rmfgen} and {\tt arfgen}. Spectra are rebinned using {\tt grppha} with a minimum of 25 counts per bin. We also use {\tt epiclccorr} to generate the background-subtracted light curves and correct for vignetting and dead-time effects.

The source was also monitored by the XMM-Newton Optical Monitor (OM) with UVW1, UVM2 and UVW2 filters during the observation period. We reduced the OM data using {\tt omichain} to produce calibrated images and source lists. Photometry was then extracted using {\tt omsource}. 


\begin{table}
    \centering
    \begin{tabular}{c|c|c}
    \hline \hline
    Obs. ID & Obs. date (MJD) & Exp. (ks) \\
    \hline
    0893810201 & 2021-12-31 (59579) & 55.5 \\
    0893810401 & 2022-01-19 (59598) & 77.6 \\
    0910190101 & 2022-06-30 (59760) & 99.1 \\
    0910190701 & 2022-07-13 (59773) & 109.1 \\
    0910190901 & 2022-07-23 (59783) & 111.6 \\
    0910191101 & 2022-08-02 (59793) & 102.9 \\
    0910191301 & 2022-08-10 (59801) & 102.9 \\
    \hline \hline
    \end{tabular}
    \caption{Summary of XMM-PN observations of SDSS J1430. Columns list the observation ID, the observation date with the corresponding MJD, and the net exposure time after background flare filtering.}
    \label{tab:pn_obsids}
\end{table}

\subsection{Swift}
SDSS J1430 was also monitored by Swift \citep{Gehrels2004ApJ...611.1005G}, with a cadence of roughly two days between 2021-11 and 2024-03. We followed standard data-reduction procedures using the {\tt xrtpipeline} task in {\tt HEASOFT} (v6.31), applying the latest calibration files to produce the XRT event lists. Source products were extracted from a circular region with a radius of 35$^{\prime\prime}$, while background products were obtained from a nearby source-free circular region of the same size. Net count rates in the 0.2--10 keV band were then calculated for each observation. 

Furthermore, the Swift Ultra-Violet/Optical Telescope (UVOT; \citealt{Roming2005SSRv..120...95R}) primarily observed the source with the UVW1 filter, with the remaining filters used only during a few initial and final epochs. We processed the UVOT data using the {\tt uvotsource} task, extracting source photometry from an 8$^{\prime\prime}$-radius circular aperture, yielding AB magnitudes calibrated to the Swift photometric system \citep{Breeveld2011AIPC.1358..373B}. The background was estimated from a 25$^{\prime\prime}$ radius circular region in a nearby source-free area. Detailed information on the Swift observations is provided in the supplementary material.

\subsection{ZTF}
The source was also extensively monitored by the Zwicky Transient Facility (ZTF; \citealt{Bellm2019PASP..131a8002B}), an optical time-domain survey utilizing the 48-inch Samuel Oschin Schmidt Telescope at Palomar Observatory. The ZTF observations provided a rich dataset with a high cadence of approximately two days. We retrieve all the public data via the NASA/IPAC Infrared Science Archive (IRSA; \citealt{ZTF2025}), and construct the light curves in the $g$ and $r$ bands using the ZTF forced PSF-fit photometry \citep{Masci2019PASP..131a8003M}. {Only non-flagged, high-quality data with a signal-to-noise ratio (S/N) greater than 3 are included in our subsequent analysis.}

\subsection{Pan-STARRS}
{The source was also observed by the Panoramic Survey Telescope and Rapid Response System 1 (Pan-STARRS1; \citealt{Chambers2016arXiv161205560C}) between February 2011 and July 2013. We retrieved the public Data Release 1 (DR1; \citealt{Chambers2017yCat.2349....0C}) photometric data and constructed the light curves in the $g$ and $r$-bands. Only non-flagged, high-quality data with a SNR$>$3 are included in our subsequent analysis.}

\subsection{DESI}
We also obtained an optical spectrum of the source from the Dark Energy Spectroscopic Instrument (DESI; \citealt{DESI2016arXiv161100036D,DESI2022AJ....164..207D}). DESI is a highly multiplexed, fiber-fed optical spectrograph mounted on the Mayall 4-meter telescope at Kitt Peak National Observatory. The target (Target ID: 39628329787067350) was observed by DESI on 2022-01-27, with a single exposure of 512.5 s. We retrieved the fully calibrated, one-dimensional spectrum from the official DESI Data Release 1 (DR1; \citealt{DESI2024AJ....168...58D}). The data were automatically processed by the standard DESI spectroscopic reduction pipeline \citep{Guy2023AJ....165..144G}, which includes instrumental signature removal, sky subtraction, fiber flat-fielding, and flux calibration. The resulting spectrum provides broad, continuous wavelength coverage from approximately 3600 to 9800 \AA\ with a spectral resolution of R$\sim$2000-5000.

\subsection{SDSS}
{We also use an archival optical spectrum of SDSS J1430 obtained by the Sloan Digital Sky Survey (SDSS; \citealt{York2000AJ....120.1579Y}) on 2005 May 4, which is the earliest optical spectrum of the source known to us. We estimate synthetic $g$ and $r$-band magnitudes by integrating the flux-calibrated spectrum over the corresponding SDSS filter wavelength ranges, assuming a constant transmission of unity within each band.}

\subsection{MATCH}

The Multi-mode Autonomous Terminal for Compatible Hybrid-optics (MATCH; \citealt{Xu2026arXiv260938858X}), a 1.0 m optical telescope owned by Wuhan University, has also been conducting high-cadence monitoring of SDSS J1430 since 2025-12-10. MATCH is located at the Lenghu astronomical observing site on the Tibetan Plateau in China, which is widely recognized as a world-class astronomical site, characterized by excellent median seeing ($\sim0.75''$) and a notably high fraction of clear nights, establishing it as one of the premier observatory sites in the Northern Hemisphere \citep{Deng2021Natur.596..353D}. The 1.0 m telescope features an alt-azimuth mount combined with a Ritchey-Chrétien (RC) Cassegrain dual-reflector optical design. Operating at a focal ratio of $f/7$, the telescope is equipped with a large-format QHY411 CMOS camera and a low-resolution spectrograph (ZWO ASI6200MMPro, which is not used in this work) at the focal plane. The detector consists of a $14,304 \times 10,748$ pixel array with a physical pixel size of $3.76\ \mu\text{m} \times 3.76\ \mu\text{m}$. This optical configuration yields a high spatial resolution with a pixel scale of $0.111^{\prime\prime}\ \text{pixel}^{-1}$ and provides a wide effective field of view (FOV) of $0.44^{\circ} \times 0.33^{\circ}$. To ensure optimal performance and minimize thermal noise during observations, the CMOS detector was actively cooled to an operating temperature of $-30^{\circ}\text{C}$. Laboratory and on-site testing have demonstrated the detector's outstanding low-noise characteristics, with a readout noise ranging from $1.0$ to $3.0\ e^-$ and an exceptionally low dark current of $0.0011\ e^-\ \text{pixel}^{-1}\ \text{s}^{-1}$ (measured at $-20^{\circ}\text{C}$). 

The high-cadence photometric observations were performed in the SDSS $g$, $r$, and $i$ bands, with a single exposure time of 60s. To accurately measure the light curve of the source and mitigate the effects of host galaxy background contamination, we employed aperture photometry on the difference image. Initially, the full width at half maximum (FWHM) of field stars in each frame was measured using the package {\tt photutils} \citep{Bradley2022zndo...7419741B} in {\tt Python}. For each filter, the epoch exhibiting the minimum FWHM (i.e., the best seeing conditions) was selected as the reference image. We then performed aperture photometry with a radius of $5^{\prime\prime}$ on these reference images. The reference magnitudes are calibrated based on the Pan-STARRS catalog \citep{Chambers2016arXiv161205560C,Flewelling2020ApJS..251....7F}. For the remaining individual science images, we subtracted the corresponding reference image to generate difference images, after spatial alignment and point-spread function (PSF) matching. Finally, we conducted aperture photometry at the source position on the difference images using the same aperture. The measured differential fluxes are subsequently mathematically combined with the reference fluxes to derive the absolute fluxes for each observation.

\section{Results} \label{sec:results}
\subsection{Light curves}
\label{subsec:lc}

\begin{figure*}[h]
    \centering
    \includegraphics[width=0.9\linewidth]{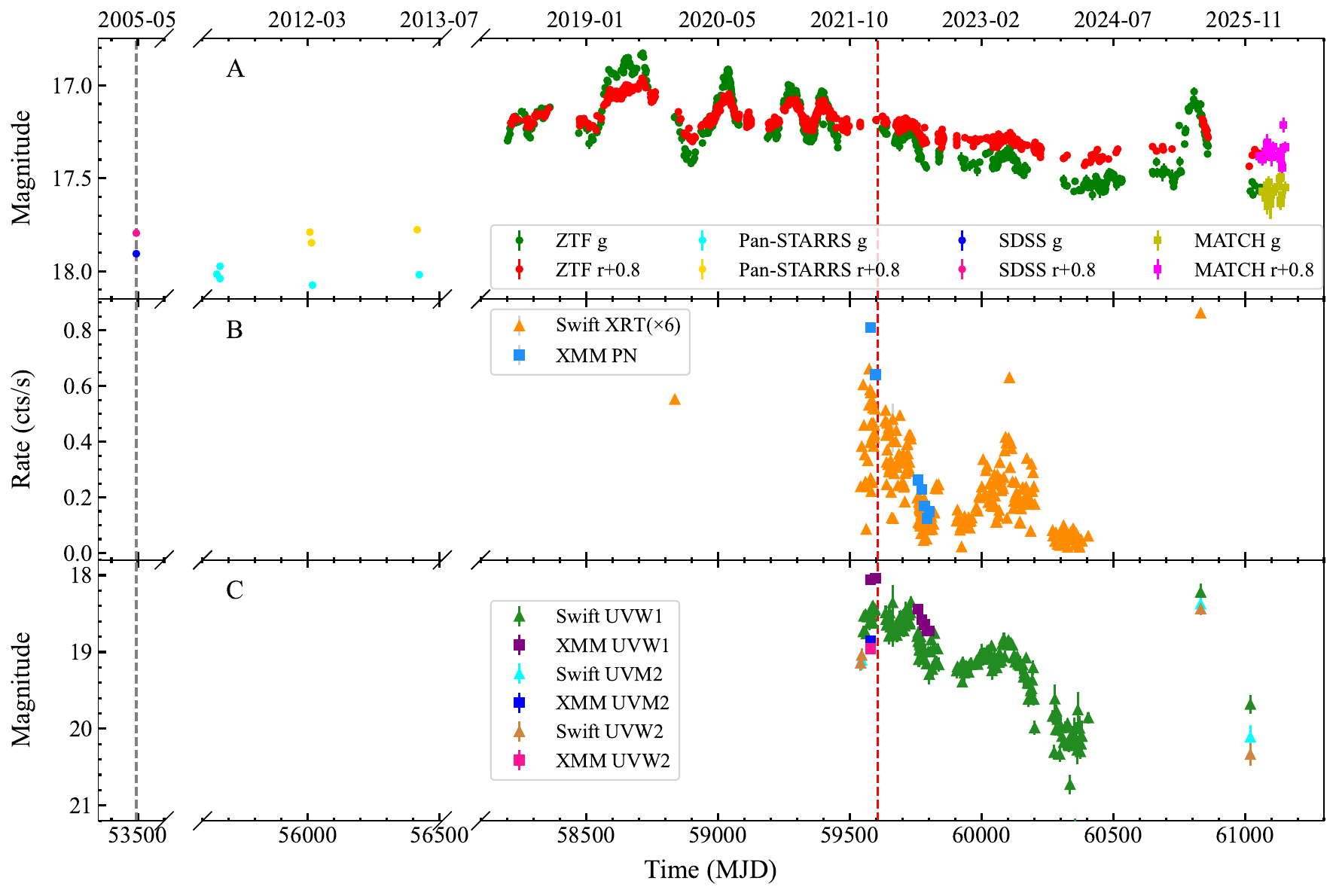}
    \caption{Broad-band light curves of SDSS J1430. (A) {Optical $g$ and $r$-band light curves from ZTF, MATCH, Pan-STARRS, and SDSS. The green/red circles, yellow/magenta squares, cyan/gold circles, and blue/pink circles represent the $g$ and $r$-band photometry from the respective surveys.} Note that the $r$-band magnitudes are shifted by 0.8 mag for clarity. (B) The X-ray light curves in 0.2-15 keV from XMM-PN (blue squares) and 0.2-10 keV from Swift-XRT (orange triangles). (C) The UV light curves from XMM-OM and Swift-UVOT. The triangles denote Swift-UVOT observations, with green, cyan, and brown representing the UVW1, UVM2, and UVW2 filters, respectively. The squares denote XMM-OM observations, with purple, navy, and pink representing the UVW1, UVW2, and UVM2 filters, respectively. Note that the gray and red dashed lines represent the spectroscopic observations of SDSS and DESI, respectively.}
    \label{fig:lc}
\end{figure*}

Using the multi-wavelength dataset assembled through the observations and data reduction procedures described in Section \ref{sec:data}, we construct the broad-band light curves of SDSS J1430, which are presented in Figure \ref{fig:lc}. {The SDSS and Pan-STARRS observations show that the source remained relatively stable between 2005 and 2013, with magnitudes of approximately $g\sim18.0$ and $r\sim17.0$. A significant brightening occurred after 2018, with the $g$ and $r$-band magnitudes peaking at $\sim16.8$ and $\sim16.2$, respectively. Thus, within at most 5 years, the source brightened by $\gtrsim1.2$ mag in $g$ and $\sim0.8$ mag in $r$.} These profound variations are consistent with the characteristic behavior of CL AGN \citep{LaMassa2015ApJ...800..144L,MacLeod2016MNRAS.457..389M,Ricci2023NatAs...7.1282R,Dotti2023MNRAS.518.4172D,Komossa2026AdSpR..77.4041K}, as confirmed by the literature and our subsequent analysis (see Section \ref{subsec:spectra type}). We hereafter refer to this brightened phase (after $\sim2018$) as the ``high state'', in contrast to the preceding quiescent ``low state'' (prior to $\sim2018$).

In the high state, the optical light curve of SDSS J1430 exhibits a long-term decline, interrupted by at least four prominent flares whose durations shorten dramatically from $\sim$400 to $\sim$100 days over three years. {Here, the flare duration is defined as the time interval between the two adjacent local minima that bracket each flare, encompassing its rising and declining phases.} The $g$ band peak slightly faded from $\sim$16.8 mag in the first flare to $\sim$17.0 mag by the fourth, while the flare amplitude progressively declined from $\sim$0.6 mag to $\sim$0.3 mag. After MJD 59600, the optical variability subsided, with the amplitude damping to only $\sim$0.1 mag. {Although the X-ray monitoring started only after the source had entered the high state}, simultaneous X-ray and UV observations (MJD 59500 -- 60400) show that the source exhibited dramatic variations. The UV flares reached amplitudes of $\sim$0.5 mag, while the X-ray emission exhibited extreme variability, with flux changes spanning up to an order of magnitude. In the later observations between MJD 60700 and MJD 60900, another prominent optical flare emerged, with a variability amplitude of $\sim$0.6 mag, comparable to the first flare observed in 2019-2020. Simultaneously, the X-ray and UV fluxes exhibited significant enhancements, exceeding those in any previous observations. We also note that the UV magnitudes from XMM-OM are systematically brighter than those from Swift-UVOT, which is due to instrumental effects between the two instruments, such as differences in the responses of the UVW1 filter, non-negligible ``red-leaks” of the filter, and the coincidence losses of source counts in the photodetector \citep{Yershov2014Ap&SS.354...97Y}.

\subsection{Color-magnitude diagram}

\begin{figure}[h]
    \centering
    \includegraphics[width=0.8\linewidth]{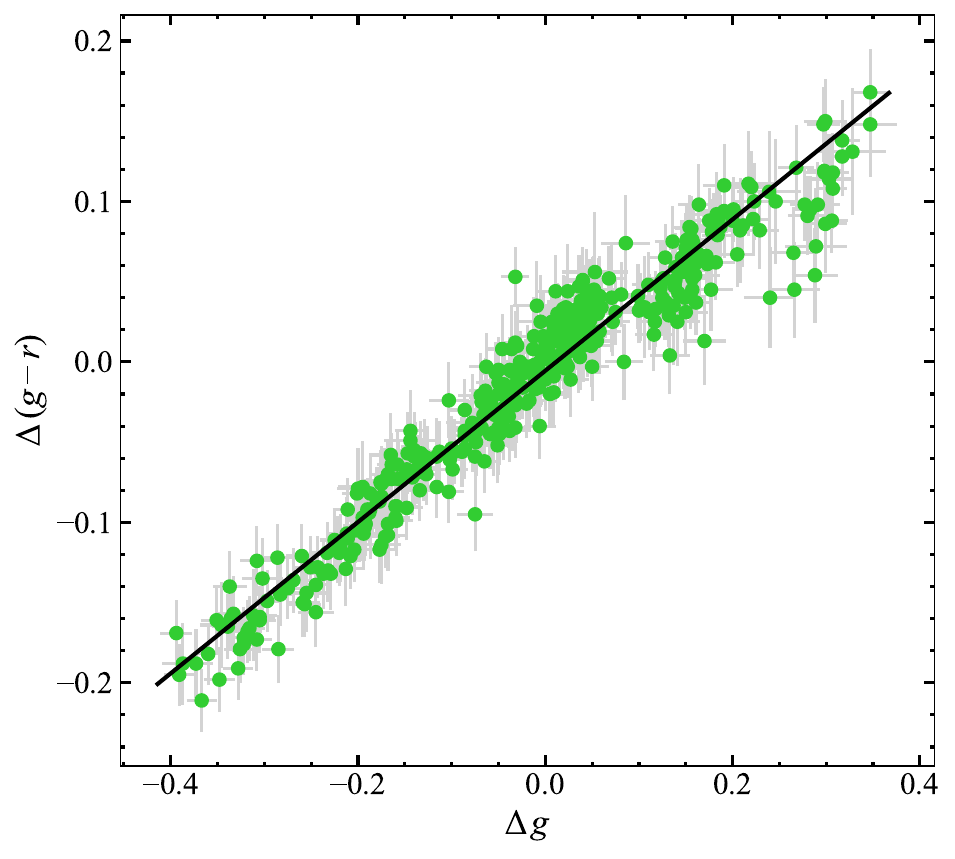}
    \caption{The color-magnitude diagram for SDSS J1430. The black line represents the best-fit linear relation.}
    \label{fig:cmd}
\end{figure}

To characterize the long-term variability properties, we examined the relation between the color variation $\Delta\,(g-r)$ and the $g$-band magnitude variation using the ZTF data, i.e., the color-magnitude diagram (CMD). Here $\Delta\,(g-r)=(g-r)-\langle g-r\rangle$ and $\Delta\,g=g-\langle g\rangle$, where $\langle g-r\rangle$ and $\langle g\rangle$ denote the median values of the magnitude of the $g-r$ and $g$-bands. The negative $\Delta\,g$ indicates that the source becomes brighter, and the negative $\Delta\,(g-r)$ corresponds to a bluer color. As shown in Figure \ref{fig:cmd}, the source exhibits a strong positive correlation between color and magnitude, with the Spearman’s coefficient $r_{\rm s}=0.96$ (significance $>99.9\%$), confirming the well-known ``bluer-when-brighter'' (BWB) trend commonly observed in unobscured AGNs \citep{Ulrich1997ARA&A..35..445U,Vanden2004ApJ...601..692V,Wilhite2005ApJ...633..638W,Schmidt2012ApJ...744..147S}. This behavior is generally attributed to an increase in the accretion disk temperature at higher accretion rates \citep{Pereyra2006ApJ...642...87P,Sakata2010ApJ...711..461S,Ruan2014ApJ...783..105R}. To quantify the color variability, we then fit the relation with a linear function:
\begin{equation}
    \Delta\,(g-r)=k\,\Delta g + b,
\end{equation}
and obtain a best-fit slope of $k=0.471\pm0.006$ with $b=-0.006\pm0.001$. This value is consistent, within the uncertainties, with the statistical results of $k=0.61\pm0.21$ for CL AGN and $k=0.53\pm0.27$ for Type 1 AGNs \citep{Wang2026ApJ...997..100W}. 

\subsection{X-ray spectra modeling}
\label{subsec:spectra fit}
\begin{figure*}[h]
    \centering
    \includegraphics[width=0.7\linewidth]{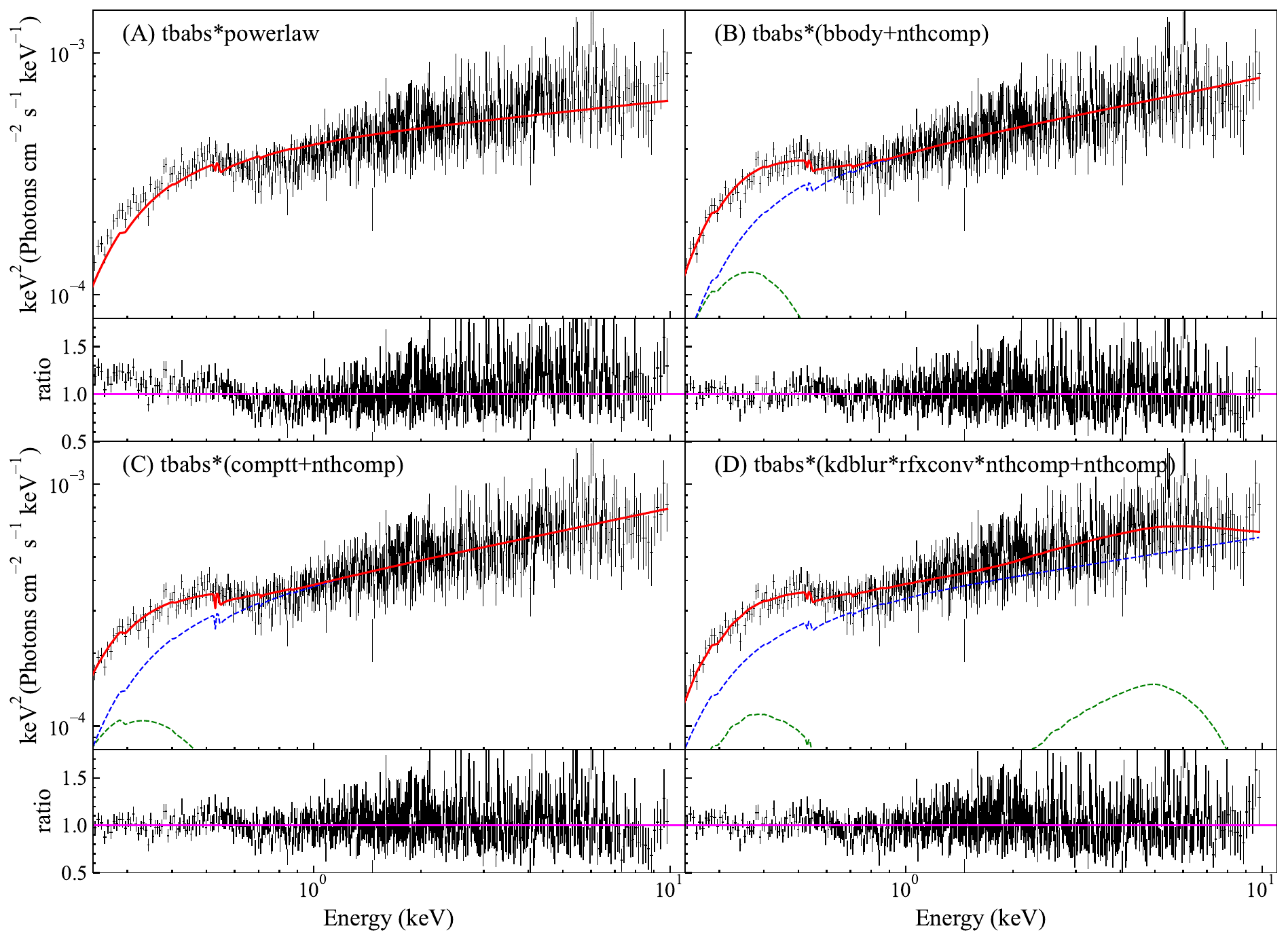}
    \caption{Four models of X-ray spectral fitting for SDSS J1430 from the first XMM-PN observation with Obs. ID 0893810201. In each panel, black points denote the unfolded XMM-PN spectrum, and the red solid line represents the best-fitting model. The lower sub-panel of each panel shows the data-to-model ratio, with black points indicating the ratio and the magenta solid line marking ratio=1. In panels (B)-(D), the blue dashed line represents the hard Comptonization component, while the green dashed line represents the blackbody component in panel (B), the soft Comptonization component in panel (C), and the reflection component in panel (D). }
    \label{fig:spec_fit}
\end{figure*}

As shown in Figure \ref{fig:lc}, the X-ray light curve reveals a striking decline, with the count rate decreasing significantly by a factor of $\sim6$ from 2021-12-31 to 2022-08-10. To investigate the temporal evolution of the X-ray spectrum of SDSS J1430, we perform the spectral fit to {all 7 epochs of XMM-PN data} in 0.2-10 keV via {\tt XSPEC} (version 12.13.0). We initially test a simple absorbed power-law model {\tt tbabs*powerlaw}. As an example, the unfolded X-ray spectrum and the best-fitting model from the first XMM-PN observation (Obs. ID: 0893810201) are shown in Figure \ref{fig:spec_fit}A, and the best-fitting parameters are listed in Table \ref{tab:spec_para}. The {\tt tbabs} model \citep{Wilms2000ApJ...542..914W}, accounting for the Galactic absorption, is fixed at the column density of $N_{\rm H} = 2.26\times 10^{20}~\rm cm^{-2}$. However, this simple model leaves significant residuals in the soft X-ray band, resulting in poor spectral fits with reduced chi-squared values of $\chi^2_{\nu}=\chi^2/\text{dof}>1.2$ across all epochs, except for MJD 59793 with $\chi^2_{\nu}=1.16$. To improve the fit, we add an additional phenomenological model, a single-temperature blackbody ({\tt bbody}) for the soft excess \citep{Ricci2017ApJS..233...17R}, and use the Comptonization model {\tt nthcomp} \citep{Zdziarski1996MNRAS.283..193Z} to describe the hard X-ray continuum arising from thermal Comptonization. During the spectral fitting, the electron temperature $kT_{\rm hot}$ in {\tt nthcomp} is fixed at 100 keV, and its seed photon temperature is tied to the temperature of the seed photons in the blackbody component. The example fitting result is presented in Figure \ref{fig:spec_fit}B, and the best-fitting parameters are listed in Table \ref{tab:spec_para}. The {\tt tbabs*(bbody+nthcomp)} model now yields statistically acceptable fits for all spectra, with the reduced chi-square $\chi^2_\nu$ ranging from 0.90 to 1.12.

{While the single blackbody provides statistically acceptable fits to the soft excess, its normalization ($\sim 5\times10^{-6}$) implies an emitting radius of only $\sim0.01\,R_{\rm g}$, which is physically implausible (see Appendix \ref{app:bbody}). This suggests that the blackbody component may provide only a phenomenological description.} Indeed, the physical origin of the AGN soft excess remains debated, with the two leading scenarios invoking Comptonization in a warm corona \citep{Magdziarz1998MNRAS.301..179M,Done2012MNRAS.420.1848D,Petrucci2013A&A...549A..73P,Rozanska2015A&A...580A..77R}, or blurred ionized reflection \citep{Crummy2006MNRAS.365.1067C}. In the first scenario, the seed photons from the accretion disk are Comptonized by a warm corona above the disk, which is optically thicker and cooler than the corona that generates the hard X-ray continuum. In the ionized reflection scenario, the emission lines produced in the disk are relativistically blurred due to the proximity to the black hole, leading to a broad, featureless pseudocontinuum in the soft X-ray band.

To model the warm corona, we replace the single blackbody with a cool, optically thick Comptonization component {\tt comptt}, which is an appropriate representation for the disk-corona geometry in \cite{Done2012MNRAS.420.1848D}. In the spectral fits, we fix the electron temperature in {\tt nthcomp} at 100 keV \citep{Done2012MNRAS.420.1848D,Jin2021ApJ...912...20J} {and its seed photon temperature is tied to the free soft photon temperature $T_0$ in {\tt comptt}. The electron temperature $kT_{\rm e}$ and optical depth $\tau$ of the warm corona in {\tt comptt} are degenerate and cannot be independently constrained by the available data. We therefore fix $\tau=20$, within the typical range of $\tau\sim10-30$ adopted for Type 1 AGN \citep{Jin2012MNRAS.420.1825J,Ballantyne2024MNRAS.530.1603B}. We also examine the effects of different $\tau$ on the fitting results (see Appendix \ref{app:diff_tau}).} This configuration yields statistically good fits for all epochs, with $\chi^2_\nu$ ranging from 0.92 to 1.16. Figure \ref{fig:spec_fit}C shows an example of the unfolded spectrum and the best-fit model, with the corresponding best-fitting parameters listed in Table \ref{tab:spec_para}. The fitting results for all epochs are presented in Appendix \ref{app:eeuf_all}. 

As an alternative to the warm corona scenario, we also explore the ionized reflection model by replacing {\tt comptt} with the reflection convolution model {\tt rfxconv} \citep{Ross2005MNRAS.358..211R,Kolehmainen2011MNRAS.416..311K}. We fix the seed photon temperature in {\tt nthcomp} at the temperature obtained from the warm corona modeling, and freeze the electron temperature in {\tt nthcomp} at 100 keV. Since the atomic features in the reflection spectrum are not clearly resolved in our data, we convolve {\tt rfxconv} with the relativistic blurring model {\tt kdblur} \citep{Laor1991ApJ...376...90L} to account for relativistic smearing. The final fitting model is {\tt tbabs*(kdblur*rfxconv*nthcomp+nthcomp)}. We set the emissivity and the inner disk radius $R_{\rm in}$ in {\tt kdblur} as the free parameters, and assume an inclination of $30^\circ$ and the iron abundance $A_{\rm Fe}=1$. The reflection model yields a statistically acceptable fit with $\chi_\nu^2=0.95\sim1.19$, and the best-fitting result is shown in Figure~\ref{fig:spec_fit}D. The best-fitting parameters are also listed in Table \ref{tab:spec_para}. We also present the fitting results for all epochs in Appendix \ref{app:eeuf_all}. 

\begin{figure}[h]
    \centering
    \includegraphics[width=0.7\linewidth]{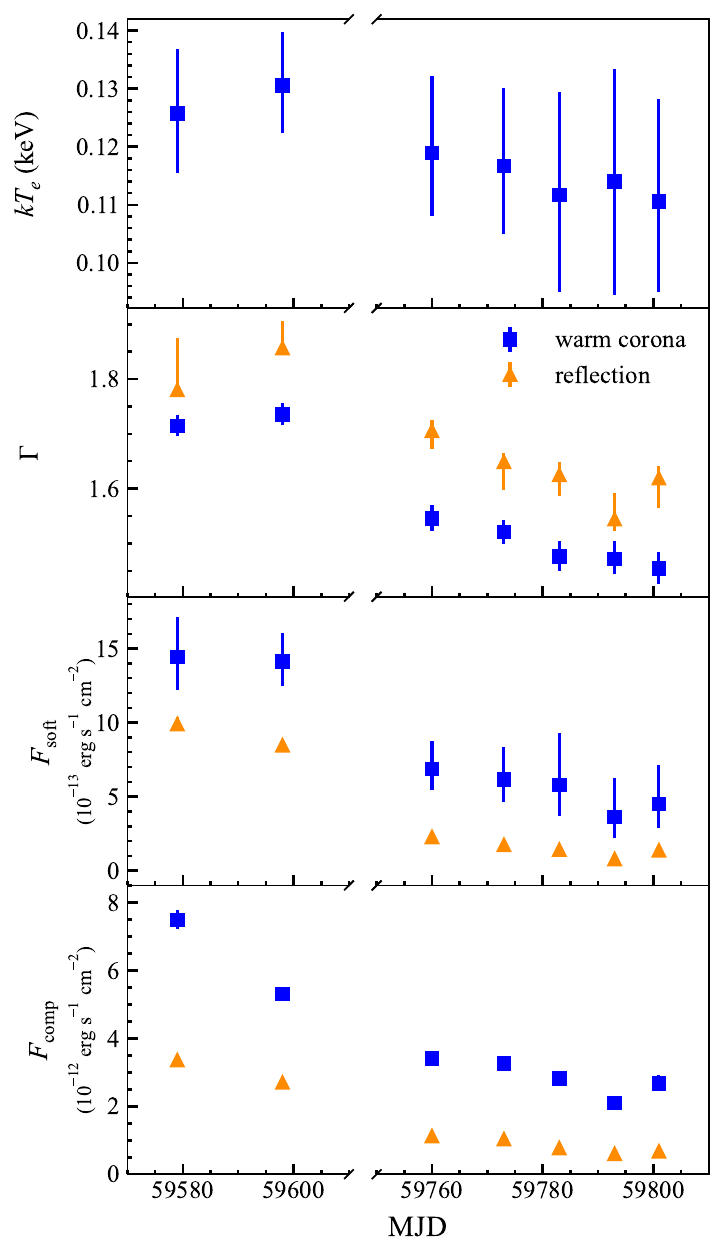}
    \caption{{Temporal evolution of the spectral parameters from the warm corona (blue squares) and blurred reflection (orange triangles) models.} From top to bottom: the electron temperature $kT_{\rm e}$ of the soft Comptonization component in the warm corona model, the photon index $\Gamma$ of the hard Comptonization component, and the unabsorbed fluxes of the soft and hard Comptonization components in the 0.01-100 keV band. All error bars represent the 1-$\sigma$ confidence level.}
    \label{fig:spec_para}
\end{figure}

{Although the warm corona and blurred reflection models both provide comparably good fits to the data, they cannot be unambiguously distinguished with the current observations, which requires additional UV observations to constrain the UV part of the spectrum, or sources with a much stronger soft X-ray excess. Given the weak Fe K$\alpha$ feature in 6-7 keV, the warm corona interpretation appears more plausible, but it remains inconclusive.} We also show the evolution of the key parameters for both models in Figure \ref{fig:spec_para}. In the warm corona model, $kT_{\rm e}$ remains nearly constant at 0.11-0.13 keV, consistent with typical AGN warm corona temperatures \citep{Gierlinski2004MNRAS.349L...7G}. {We find that the major changes in the spectral shape occur between the first two epochs and the subsequent five epochs}, as the photon index $\Gamma$ of the hard Comptonization component in the two models shows similar decreasing trends, although $\Gamma$ is systematically lower by $\sim$0.1 in the warm corona model. Furthermore, we use the {\tt cflux} task to estimate the unabsorbed fluxes of the soft component ($F_{\rm soft}$) and the hard component ($F_{\rm comp}$) in 0.01-100 keV for both models. Both models show consistent temporal evolution in these components, while the warm corona model yields systematically higher fluxes than the reflection model.

\begin{table*}
    \centering
    \begin{tabular}{llcc}
    
    \hline\hline
    Model/$\chi_\nu^2$ & Component & parameter & value \\
    \hline
    {\tt tbabs*powerlaw} & & & \\
    $\chi_\nu^2=1.53$ & {\tt tbabs} & $N_{\rm H}~(10^{20}~\rm cm^{-2})$ & 2.26 (fixed) \\
     & {\tt powerlaw} & $\Gamma$ & $1.84^{+0.01}_{-0.01}$ \\
    \hline
    {\tt tbabs*(bbody+nthcomp)} & & & \\
    $\chi_\nu^2=1.08$ & {\tt tbabs} & $N_{\rm H}~(10^{20}~\rm cm^{-2})$ & 2.26 (fixed) \\
     & {\tt bbody} & $kT~(\rm eV)$ & $72.77^{+4.25}_{-4.34}$ \\
     & {\tt bbody} & norm & $5.42^{+0.48}_{-0.45}\times10^{-6}$ \\
     & {\tt nthcomp} & $\Gamma$ & $1.72^{+0.02}_{-0.02}$ \\
     & {\tt nthcomp} & $kT_{\rm seed}~(\rm eV)$ & tied to $kT$ \\
     & {\tt nthcomp} & $kT_{\rm hot}~(\rm keV)$ & 100 (fixed) \\
     & {\tt nthcomp} & norm & $4.02^{+0.06}_{-0.07}\times10^{-4}$ \\
    \hline
    {\tt tbabs*(comptt+nthcomp)} & & & \\
    $\chi_\nu^2=1.11$ & {\tt tbabs} & $N_{\rm H}~(10^{20}~\rm cm^{-2})$ & 2.26 (fixed) \\
     & {\tt comptt} & $T_0~(\rm eV)$ & $20.89^{+17.40}_{-0.83}$ \\
     & {\tt comptt} & $kT_{\rm e}~(\rm keV)$ & $0.13^{+0.01}_{-0.01}$ \\
     & {\tt comptt} & $\tau$ & 20 (fixed) \\
     & {\tt comptt} & norm & $0.096^{+0.003}_{-0.062}$ \\
     & {\tt nthcomp} & $\Gamma$ & $1.71^{+0.02}_{-0.02}$ \\
     & {\tt nthcomp} & $kT_{\rm seed}~(\rm keV)$ & tied to $T_0$ \\
     & {\tt nthcomp} & $kT_{\rm hot}~(\rm keV)$ & 100 (fixed) \\
     & {\tt nthcomp} & norm & $3.99^{+0.08}_{-0.07}\times10^{-4}$ \\
    \hline
    {\tt tbabs*(kdblur*rfxconv*nthcomp+nthcomp)} & & & \\
    $\chi_\nu^2=1.13$ & {\tt tbabs} & $N_{\rm H}~(10^{20}~\rm cm^{-2})$ & 2.26 (fixed) \\
     & {\tt kdblur} & Index & $9.58^{+0.16}_{-3.17}$ \\
     & {\tt kdblur} & $R_{\rm in}~({R_{\rm g}})$ & $1.51^{+0.15}_{-0.01}$ \\
     & {\tt kdblur} & $R_{\rm out}~({R_{\rm g}})$ & 100 (fixed) \\
     & {\tt kdblur} & $\theta~(\rm degree)$ & 30 (fixed) \\
     & {\tt rfxconv} & $\rm Rel_{refl}$ & $-2.25^{+0.29}_{-1.07}$ \\
     & {\tt rfxconv} & $A_{\rm Fe}$ & 1 (fixed) \\
     & {\tt rfxconv} & $\log \xi\,(\rm erg\,cm\,s^{-1})$ & $1.97^{+0.03}_{-0.97}$ \\
     & {\tt nthcomp} & $\Gamma$ & $1.78^{+0.09}_{-0.01}$ \\
     & {\tt nthcomp} & $kT_{\rm seed}~(\rm keV)$ & fixed at $T_0$ \\
     & {\tt nthcomp} & $kT_{\rm hot}~(\rm keV)$ & 100 (fixed) \\
     & {\tt nthcomp} & norm & $3.55^{+0.47}_{-0.01}\times10^{-4}$ \\     
    \hline\hline
    
    \end{tabular}
    \caption{The best-fitting values of the parameters in the four spectral fitting methods for SDSS J1430 from the first XMM-PN observation with Obs. ID 0893810201. The upper and lower limits give the $1\sigma$ confidence range. 
    }
    \label{tab:spec_para}
\end{table*}

\subsection{X-ray timing analysis}
\label{subsec:timing}
As demonstrated by our spectral modeling, the warm corona component is primarily confined to energies below 0.5 keV. Motivated by this, we employ timing analysis in this section to investigate fast variability and potential time lags between the soft and hard X-rays.

We extracted light curves in the soft X-ray (0.2–0.5 keV) and hard X-ray (0.5–10 keV) bands with a time resolution of 1 s. The lag-frequency spectra were computed using the open-source Python package {\tt stingray} \citep{Huppenkothen2019ApJ...881...39H}. By convention, a positive lag denotes that the hard X-ray variations lag behind the soft X-rays (hard lag), whereas a negative lag indicates the opposite (soft lag).

\begin{figure}[h]
    \centering
    \includegraphics[width=1\linewidth]{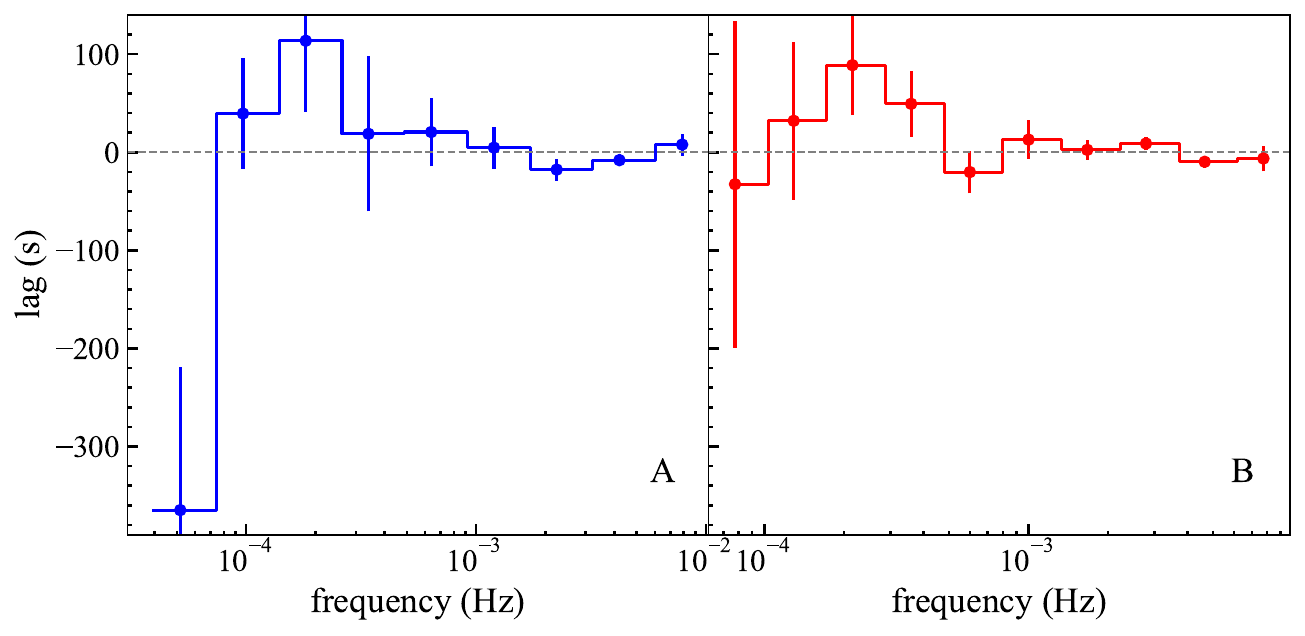}
    \caption{The time lags between 0.2--0.5 keV and 0.5--10 keV as a function of temporal frequency. Positive lags indicate that the hard band lags behind the soft band. (A) Lag-frequency spectrum derived from the brightest observation. (B) Lag-frequency spectrum obtained by combining the five fainter observations to improve the SNR. The gray dashed lines represent lag=0.}
    \label{fig:lag-freq}
\end{figure}

We first focus on the observation with the highest count rate {(Obs. ID 0893810401, 2022-01-19)} to maximize the signal-to-noise ratio (SNR). The resulting lag-frequency spectrum is shown in Figure \ref{fig:lag-freq}A. We detect a low-frequency hard lag in the frequency range of roughly $[1.5-2.5] \times 10^{-4}$ Hz, with an amplitude of $\tau = 114 \pm 72\,\rm s$. At higher frequencies ($[1.5-3.0] \times 10^{-3}$ Hz), the lag becomes negative, showing a weak soft lag with an amplitude of $\tau = -17.6 \pm 11.2$ s. To examine the timing properties during the fading phase, we also merged the light curves of the last five fainter observations to compensate for their low count rates in individual observations. The combined lag-frequency spectrum in the faint state is presented in Figure \ref{fig:lag-freq}B. Remarkably, a significant hard lag is again detected at low frequencies, $[2.8-4.8] \times 10^{-4}$ Hz, with an amplitude of $\tau = 65\pm 36$ s. This value is consistent with that observed during the brightest epoch to within the $1\sigma$ statistical uncertainties. Similarly, at higher frequencies $[3.2-6.0] \times 10^{-3}$ Hz, we observe a weak negative trend with an amplitude of $\tau = -8.8 \pm 6.7$ s. Although the soft lag amplitudes obtained from both the brightest epoch and the merged faint epochs are remarkably consistent within error bars, their absolute values are negligible, consistent with zero. Therefore, we conservatively suggest that no statistically significant soft lag is detected in this source.

The detection of a low-frequency hard lag and a high-frequency soft lag (though it is extremely weak in this work) is a well-established observational signature seen in both accreting supermassive black holes in AGNs \citep{Fabian2009Natur.459..540F,Kara2016Natur.535..388K} and stellar-mass black holes in X-ray binaries \citep{You2026NatCo..17.2860Y}. The high-frequency soft lag is commonly attributed to relativistic X-ray reverberation \citep{Uttley2014A&ARv..22...72U}, whereas the low-frequency hard lag is generally interpreted within the framework of mass accretion rate fluctuations propagating inward through the accretion disk \citep{Kotov2001MNRAS.327..799K,Arevalo2006MNRAS.367..801A}. As the fluctuations move toward the central black hole, they energize the soft-emitting outer regions before reaching the hotter, hard X-ray-emitting inner corona. Based on this propagation scenario, by comparing the lag profiles between the two observational states of our source, we find that the characteristic amplitude and frequency of the hard lag remain constant within uncertainties. Although the intrinsic X-ray luminosity decreased significantly between these two epochs, the stability of the hard lag indicates that the geometry of the inner disk-corona system changed only marginally during the fading phase.

\begin{figure}[h]
    \centering
    \includegraphics[width=1\linewidth]{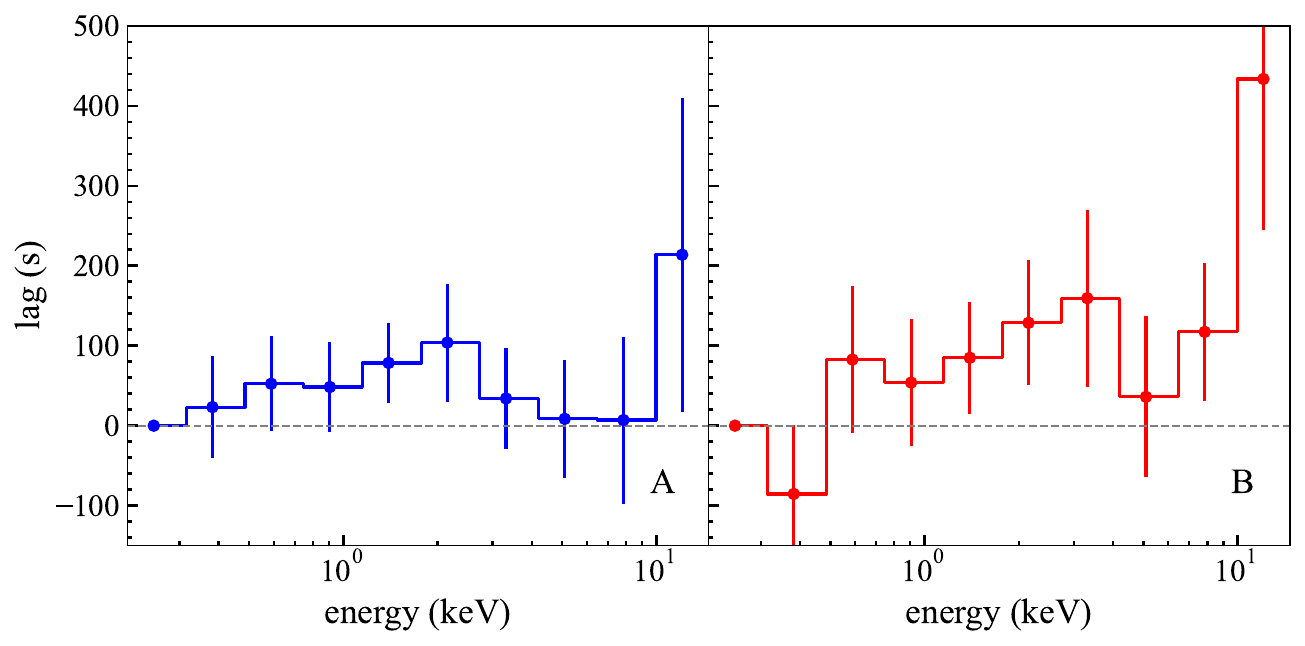}
    \caption{The lag-energy spectrum in the low frequency range $[0.8-5.0] \times 10^{-4}$ Hz for the brightest observation (A), and $[1.5-4.0] \times 10^{-3}$ Hz for the combined fainter observations (B). Positive lags indicate that the hard band lags behind the soft band. The gray dashed lines represent lag=0.}
    \label{fig:lag-energy}
\end{figure}

Furthermore, we compute the low-frequency lag spectrum for the brightest observation in $[0.8-5.0] \times 10^{-4}$ Hz and the combined fainter observations in $[1.5-4.0] \times 10^{-3}$ Hz. As illustrated in Figure \ref{fig:lag-energy}, the lag-energy spectrum exhibits an energy dependence. Below $\sim$3 keV, the time lag increases roughly linearly with energy on a logarithmic scale, reaching a maximum delay of $\sim100$ s at 2-3 keV. This is in agreement with the mass accretion rate fluctuations propagating inward through a corona, where harder X-rays are emitted from the hotter, more compact, and inner regions. Interestingly, the lag spectrum flattens or even decreases above $\sim$3 keV, though the uncertainties at higher energies are substantial due to limited photon statistics.

\section{Discussion} \label{sec:discussion}

In the high state, SDSS J1430 exhibited striking optical variability characterized by rapidly decaying periods with progressively decreasing amplitudes. Toward the end of the oscillation phase, X-ray spectral analysis shows that the spectra harden as the source fades, while X-ray timing analysis indicates a stable disk-corona geometry throughout the luminosity decline. In this section, we will discuss the spectral type transition, the $\Gamma$-Eddington ratio relation, and the observed soft X-ray excess. We will also explore the possible physical mechanism responsible for the observed oscillating optical light curve.

\subsection{Spectral type transition}
\label{subsec:spectra type}
\begin{figure*}
    \centering
    \includegraphics[width=0.8\linewidth]{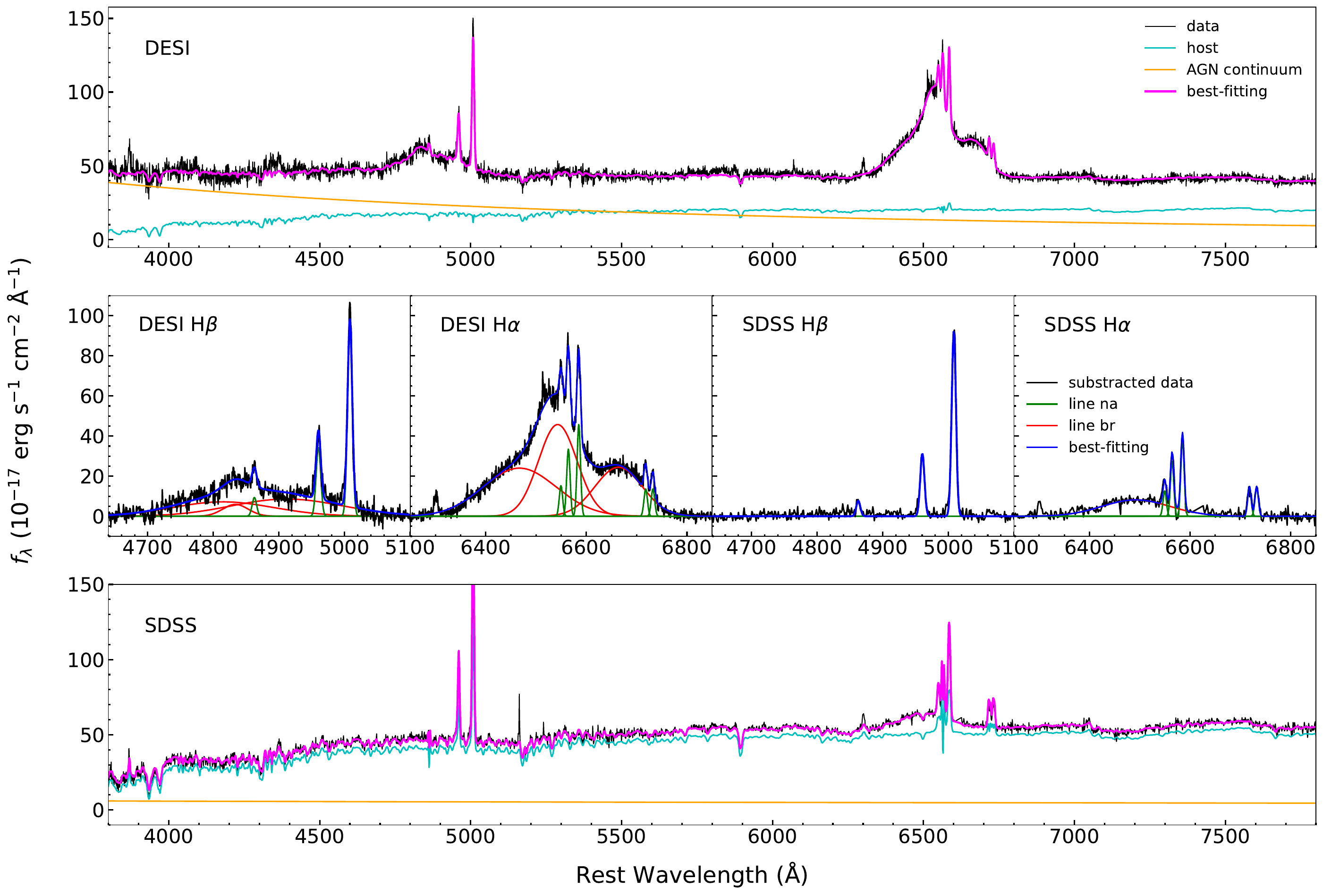}
    \caption{Optical spectral fitting of SDSS J1430. {The upper and lower panels show the full rest-frame spectra from DESI and SDSS observations, respectively.} The black points represent the de-redshifted data. The magenta line represents the best-fitting model. The cyan and orange lines show the host galaxy component and the AGN power-law continuum, respectively. The middle panels show zoom-in views of the host-subtracted H$\beta$ and H$\alpha$ regions. The black points are the host-subtracted data. The blue line shows the best-fitting model. The red and green lines represent the broad and narrow emission line components, respectively.}
    \label{fig:opt_spec}
\end{figure*}

To investigate the spectral type transition of SDSS J1430, we analyze both its archival SDSS spectrum observed on 2005-05-04 and the DESI spectrum obtained on 2022-01-27. We perform the spectral fitting using the publicly available code {\tt PyQSOFit} \citep{Guo2018ascl.soft09008G,Shen2019ApJS..241...34S,Ren2024ApJ...974..153R}. We first correct the spectrum for Galactic dust extinction using the dust reddening maps from \cite{Schlegel1998ApJ...500..525S} via the {\tt sfdmap2} package. To isolate the pure AGN emission, the de-reddened spectrum is decomposed to remove the contribution from the host galaxy starlight. The host galaxy subtraction is performed using the Principal Component Analysis (PCA) method, utilizing the host galaxy templates \citep{Yip2004AJ....128..585Y,Yip2004AJ....128.2603Y} adopted from the Penalized PiXel-Fitting ({\tt pPXF}; \citealt{Cappellari2023}). Furthermore, comprehensive optical and UV Fe II templates included in {\tt pPXF}, along with an intrinsic power-law component, are incorporated into the continuum model. Emission lines are then fitted using multiple Gaussian components. For the DESI spectrum, three Gaussians are used to model the broad H$\alpha$ and H$\beta$ profiles, accounting for their prominent blue and red wings. In contrast, the broad H$\alpha$ in the SDSS spectrum is modeled with a single Gaussian, while no broad H$\beta$ component is required. The narrow components of H$\alpha$, H$\beta$, and other emission lines are each modeled with a single Gaussian. We adopt an FWHM threshold of $1200~\rm km~s^{-1}$ to distinguish broad and narrow emission-line components. The fitting results are shown in Figure \ref{fig:opt_spec}.


The emission line fitting results reveal a dramatic transition in the broad-line region (BLR) of SDSS J1430. Based on the archival SDSS spectrum, the source was classified as a Seyfert 1.9 galaxy, characterized by the presence of broad H$\alpha$ but the absence of broad H$\beta$ \citep{Winkler1992MNRAS.257..677W}. However, the DESI spectrum shows a distinct, robust, broad H$\beta$ component emerging alongside the prominent broad H$\alpha$ with a red wing. With the blue wing observed both in the DESI and previous SDSS spectra, the asymmetric H$\alpha$ profile shows the double-peaked broad line structure, which is common among optically variable AGNs \citep{Ward2024ApJ...961..172W}. Similar H$\alpha$ profiles have also been reported from observations with the Lijiang 2.4-m telescope \citep{Jiang2022arXiv220111633J} and the Seimei Telescope \citep{Hoshi2024PASJ...76..103H}. The measured flux ratio of H$\beta$/[O III]$\lambda$5007=4.6 firmly classifies SDSS J1430 as a Seyfert 1.2 galaxy in the current epoch \citep{Winkler1992MNRAS.257..677W}. This evolution from Seyfert 1.9 to Seyfert 1.2 identifies SDSS J1430 as a ``turn-on'' CL AGN \citep{MacLeod2016MNRAS.457..389M,Ricci2023NatAs...7.1282R,Dotti2023MNRAS.518.4172D,Guo2025ApJS..278...28G,Komossa2026AdSpR..77.4041K,Wang2026ApJ..1002...85W}. {Furthermore, we estimate the single-epoch BH mass using the total luminosity and the overall FWHM of the broad emission line profiles obtained by combining the multiple Gaussian components.} {The resulting BH masses are $M_{\rm BH}=5.3\pm1.3\times10^{7}~\rm M_{\odot}$ from the SDSS broad H$\alpha$, and $M_{\rm BH}=1.15\pm0.33\times10^{8}~\rm M_{\odot}$ and $1.95\pm0.08\times10^{8}~\rm M_{\odot}$ from the DESI broad H$\alpha$ and H$\beta$, respectively \citep{Greene2005ApJ...630..122G}. The SDSS-based estimate is close to the previously reported value of $M_{\rm BH}\sim4\times10^{7}~\rm M_\odot$, which was also derived from the broad H$\alpha$ line in the SDSS spectrum \citep{Jiang2022arXiv220111633J}. The DESI-based estimates are broadly consistent with the value of $\sim2\times10^{8}~\rm M_\odot$ inferred from the host-galaxy scaling relations \citep{Jiang2022arXiv220111633J}. The discrepancy between the SDSS- and DESI-based estimates may be partly attributed to the different spectral states of the source, which exhibit distinct broad line profiles and widths, thus leading to substantial differences in the inferred BH masses.}


\subsection{$\Gamma$-$\lambda_{\rm Edd}$ relation}

{The relation between $\Gamma$ and the Eddington ratio $\lambda_{\rm Edd}$ is closely linked to the accretion physics and disk-corona coupling in both AGN and black hole X-ray binaries (BHXRBs) (e.g., \citealt{Shemmer2006ApJ...646L..29S,Brightman2013MNRAS.433.2485B,Yang2015MNRAS.447.1692Y,Trakhtenbrot2017MNRAS.470..800T}). At high accretion rates, $\Gamma$ is positively correlated with $\lambda_{\rm Edd}$. Several studies have reported that this trend reverses at lower accretion rates, which produces a V-shaped $\Gamma$--$\lambda_{\rm Edd}$ relation with a turnover at $\log\lambda_{\rm Edd}\sim-2$ to $-2.5$ (e.g., \citealt{Gu2009MNRAS.399..349G,She2018ApJ...859..152S,Diaz2023A&A...669A.114D}). For instance, \citet{She2018ApJ...859..152S} found a break at $\log\lambda_{\rm Edd}\sim-2.5$ in a sample of nearby AGN observed with \textit{Chandra}, and \citet{Diaz2023A&A...669A.114D} reported a similar break at $\log\lambda_{\rm Edd}\sim-2.39$ in low-luminosity AGN. This turnover may mark a transition between distinct accretion regimes, analogous to the hard-to-soft state transitions in BHXRBs (e.g., \citealt{Remillard2006ARA&A..44...49R,Gu2009MNRAS.399..349G,Jana2022MNRAS.512.5942J}). The same analogy has been invoked for CL AGN, for which both theoretical models and observational studies suggest accretion-state transitions similar to those in BHXRBs (e.g., \citealt{Noda2018MNRAS.480.3898N,Ross2018MNRAS.480.4468R,Ruan2019ApJ...883...76R,Ai2020ApJ...890L..29A,Yan2020ApJ...889L..18Y,Hagen2024MNRAS.534.2803H,Kang2025MNRAS.538..121K}).}

\begin{figure}
    \centering
    \includegraphics[width=0.9\linewidth]{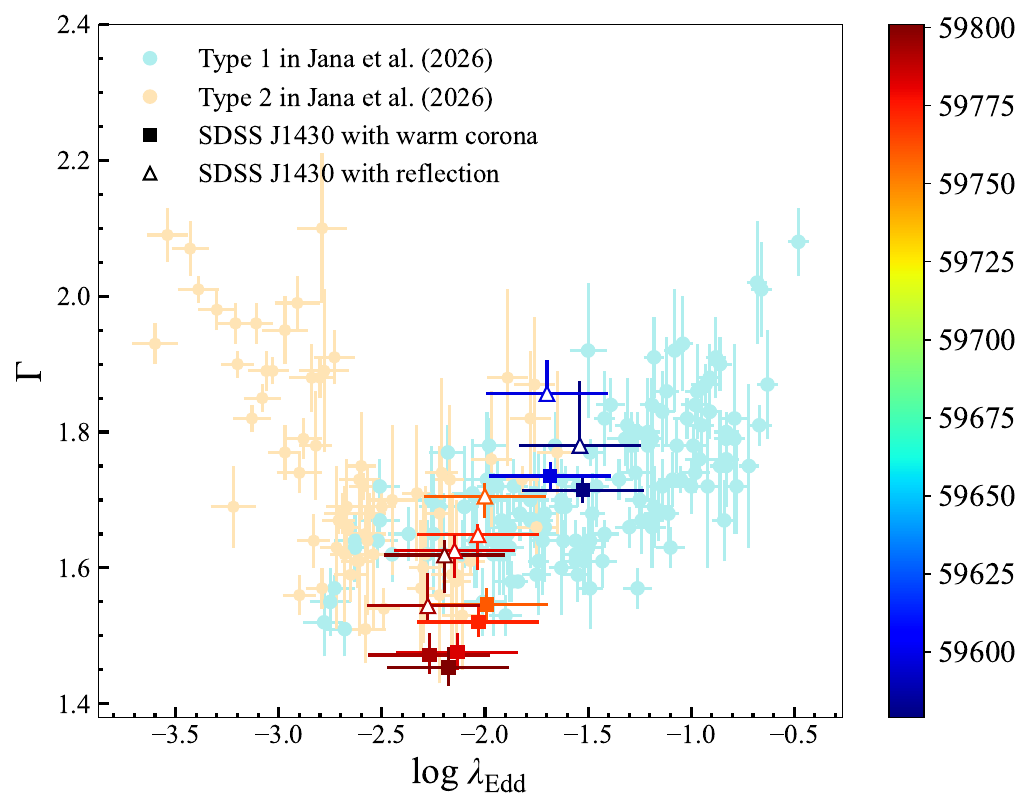}
    \caption{The relation between $\Gamma$ and Eddington ratio $\lambda_{\rm Edd}$. The filled squares represent the results for SDSS J1430 obtained with the warm corona model, {while the hollow triangles represent those obtained with the reflection model}, with color-coded by observation epoch. The filled circles show the comparison sample of \cite{Jana2026A&A...707A.213J}, with blue and orange denoting Type 1 and Type 2 states, respectively.}
    \label{fig:gamma}
\end{figure}

{We also plot the relation between $\Gamma$ and the Eddington ratio in Figure \ref{fig:gamma}. The bolometric luminosity $L_{\rm bol}$ is estimated using
\begin{equation}
    L_{\rm bol}=K_{\rm X}\,L_{\rm X},
\end{equation}
where $L_{\rm X}$ is the hard X-ray luminosity in 2-10 keV and $K_{\rm X}$ is the hard X-ray bolometric correction. Following \citet{Duras2020A&A...636A..73D}, we adopt the luminosity-dependent correction:
\begin{equation}
K_{\rm X}=a\left[1+\left(\frac{\log(L_{\rm bol}/L_{\odot})}{b}\right)^c\right],
\end{equation}
where $a=12.76$, $b=12.15$, and $c=18.78$ are the best-fitting coefficients for Type 1 AGNs. Since $K_{\rm X}$ depends on $L_{\rm bol}$, we solve the two equations simultaneously through an iterative procedure to obtain $L_{\rm bol}$. This yields $K_{\rm X}=13\sim15$ and $L_{\rm bol}=8.3\sim37.1\times10^{43}\,\rm erg\,s^{-1}$. We then calculate the Eddington ratio as $\lambda_{\rm Edd}= L_{\rm bol}/L_{\rm Edd}$, where $L_{\rm Edd}$ is the Eddington luminosity, given by $L_{\rm Edd}= 1.26\times10^{38}\left(M_{\rm BH}/ M_{\odot}\right)\,{\rm erg\,s^{-1}}$. We adopt the black hole mass derived from the broad lines in Section \ref{subsec:spectra type}. }

{We find that both spectral models follow the general trend of softer X-ray spectra at higher Eddington ratios observed in Type 1 AGNs \citep{Jana2026A&A...707A.213J}. We also find SDSS J1430 lies in the low Eddington ratio regime of Type 1 AGNs, placing it near the boundary between Type 1 and Type 2 AGNs in the $\Gamma-\lambda_{\rm Edd}$ diagram.}

\subsection{Weak soft excess}
\label{subsec:weak SE}

\begin{figure}
    \centering
    \includegraphics[width=0.9\linewidth]{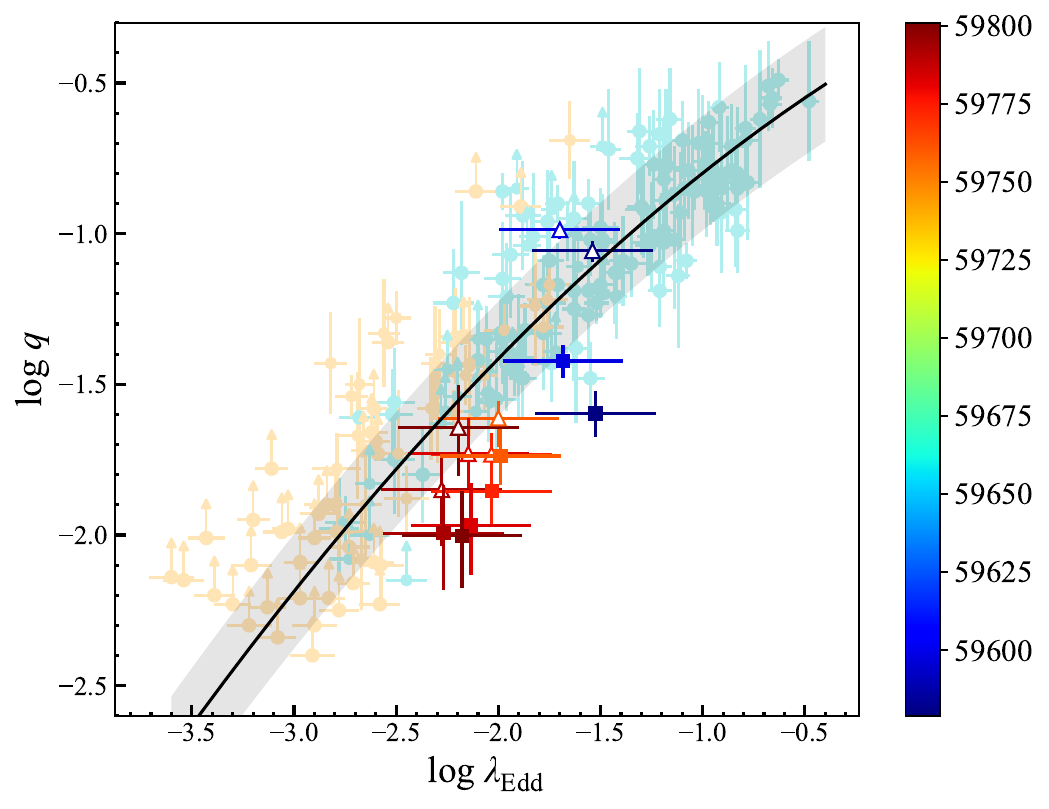}
    \caption{Soft excess strength $q$ as a function of Eddington ratio $\lambda_{\rm Edd}$. The black line and gray shaded region show the best-fitting relation and its $1\sigma$ scatter from \cite{Jana2026A&A...707A.213J}. Upward-pointing error bars indicate upper limits. Other symbols are the same as those in Figure \ref{fig:gamma}.}
    \label{fig:SE}
\end{figure}

To further quantify the relative strength of the soft excess, we calculate the soft excess strength $q$, defined as the ratio of the soft component in 0.5-2 keV to the hard X-ray continuum in 2-10 keV \citep{Jana2026A&A...707A.213J}. {The corresponding uncertainties are propagated from those of the two component fluxes.} As shown in Figure \ref{fig:SE}, {both the warm corona (circles) and blurred reflection (triangles) models yield a weak soft excess,  with $q<0.04$ for most observations, except for the first two epochs fitted with the reflection model. These values place SDSS J1430 in the low regime of previous samples \citep{Bianchi2009A&A...495..421B,Jin2012MNRAS.420.1825J,Boissay2016A&A...588A..70B,Jana2026A&A...707A.213J}. Meanwhile, the reflection model systematically gives higher $q$ values than the warm corona model. This difference arises because the warm corona emission is more concentrated at softer X-ray energies than the reflected emission, which results in a lower inferred soft component flux than in the reflection model and hence a lower $q$.}

The physical origin of the weak soft excess is closely related to the accretion rate, as previous studies have found a positive correlation between soft excess strength and Eddington ratio \citep{Boissay2016A&A...588A..70B,Jana2026A&A...707A.213J}. As shown in Figure \ref{fig:SE}, both models broadly follow the positive $q$-$\lambda_{\rm Edd}$ relation established for Type 1 and Type 2 AGNs by \citet{Jana2026A&A...707A.213J}, with both $q$ and $\lambda_{\rm Edd}$ declining monotonically along the observing sequence (color-coded by time). {We also find that the $q$ values obtained from both models generally lie below the reference relation at a given $\lambda_{\rm Edd}$, except for the first two epochs fitted with the blurred reflection model, indicating a weaker soft excess than typically observed in Type 1 AGNs. The deviation is more pronounced for the warm corona model, suggesting an even weaker soft excess compared to that inferred from the blurred reflection model.}

A striking analog can be drawn with the typical CL AGN Mrk 1018 \citep{Husemann2016A&A...593L...9H,Lyu2021MNRAS.506.4188L}. During its dramatic fading from a Type 1 to a Type 1.9 state, the Eddington ratio decreased from $\sim10^{-1.14}$ to $\sim10^{-2.27}$, accompanied by the complete disappearance of the soft excess as the accretion rate dropped \citep{Husemann2016A&A...593L...9H,Noda2018MNRAS.480.3898N}. Such a ``vanishing'' soft excess has also been reported in other CL AGNs (e.g. \citealt{Ghosh2022ApJ...937...31G}). 
This comparative evidence strongly suggests that the observed weak soft excess in SDSS J1430 is inherently driven by its low accretion rate.

\subsection{A Shrinking Unstable Zone in the Accretion Disk: Explaining the Decaying Periods and Damping Amplitudes}
\label{subsec:RPI}
During the high state, SDSS J1430 exhibits at least four prominent flares. The durations of the flares decrease remarkably from $\sim$400 to $\sim$100 days over three years, accompanied by a progressive damping in flare amplitude. Similar behavior is also observed in the recently discovered black hole X-ray binary Swift J1727.8-1613, where the flare timescale decreases from $\sim$6.42 days to $\sim$3 days, accompanied by a gradual decline in the flare amplitude, and is suggested to be related to the thermal-viscous disk instability \citep{He2026ApJ..1006..177H}. This motivates us to interpret this striking variability behavior in SDSS J1430 through the disk instability.

\cite{Sniegowska2020A&A...641A.167S} proposed a radiation pressure instability (RPI) model to explain the semi-periodic, repeating outbursts observed in CL AGNs. Later, this model was further developed by incorporating a large-scale magnetic field, which effectively shortens the outburst period \citep{Pan2021ApJ...910...97P}. The model is motivated by the fact that a standard geometrically thin, optically thick accretion disk \citep{Shakura1973A&A....24..337S} becomes unstable in its innermost region when radiation pressure dominates over gas pressure \citep{Pringle1973A&A....29..179P,Lightman1974ApJ...187L...1L,Shakura1976MNRAS.175..613S}. In sources accreting at a few percent of the Eddington rate, the inner accretion flow transitions from a cold, gas-dominated standard disk to a hot, optically thin advection-dominated accretion flow (ADAF) \citep{Narayan1994ApJ...428L..13N}. The key insight of \cite{Sniegowska2020A&A...641A.167S} is that a narrow unstable zone exists precisely at the transition radius between these two regimes, where radiation pressure instability operates on a timescale that is significantly shorter than the viscous timescale of the Shakura-Sunyaev disk. The instability timescale depends on the radial scale of the zone:
\begin{equation}
    \tau = \tau_{\rm visc}\cdot \frac{\Delta R_{\rm u}}{R_{\rm u}}~(\Delta R_{\rm u}\ll R_{\rm u})
\end{equation}
where $\Delta R_{\rm  u}$ is the radial width of the unstable zone, $R_{\rm u}$ stands for the location of the zone, $\tau_{\rm visc}$ represents the viscous timescale at $R_{\rm u}$. This dramatically shortens the timescale of variation from hundreds of years to a few years and simultaneously reduces the amount of variable radiation flux by the same factor. The resulting variability is expected to be most prominent in X-rays, which irradiate the outer disk and drive correlated variation in the optical band. However, as this remains a simplified toy model that does not account for the detailed disk structure (such as proper vertical structure, opacity, convection, \citealt{Sniegowska2020A&A...641A.167S}), we aim to test whether the RPI model roughly produces the shortening variability timescales and damping of flare amplitude observed in SDSS J1430.

\begin{figure}
    \centering
    \includegraphics[width=1\linewidth]{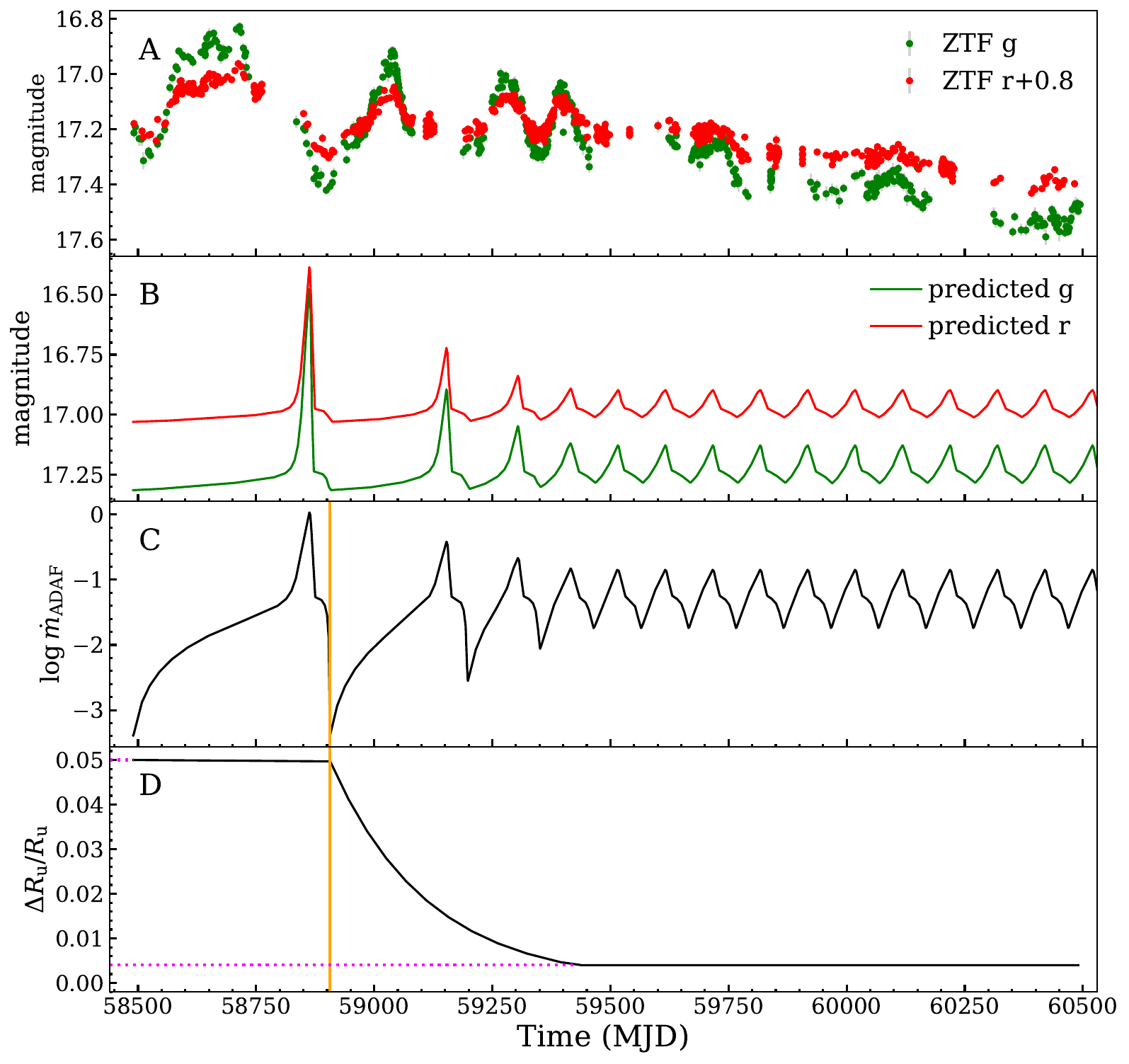}
    \caption{Results of the radiation pressure instability model for SDSS J1430. (A) Observed $g$- (green) and $r$-band (red) light curves from ZTF. Note that the $r$-band magnitudes are shifted by 0.8 mag for visualization purposes. Note that the $r$-band magnitudes are shifted by 0.8 mag for clarity. (B) Predicted $g$- and $r$-band light curves in the observer frame. (C) Predicted evolution of the accretion rate of ADAF modulated by the radiation pressure instability. (D) Temporal evolution of $\Delta R_{\rm u}$, where the magenta dotted lines mark the initial and final values of the evolution. Note that the orange lines represent the time $\Delta R_{\rm u}$ starts to evolve. }
    \label{fig:rim}
\end{figure}

The model parameters include the BH mass $M_{\rm BH}$, the external accretion disk rate $\dot{m}_{\rm disk}$, the viscosity parameter $\alpha$, the location $R_{\rm u}$, and the radial width $\Delta R_{\rm u}$ of the unstable zone. For a given set of parameters, the model self-consistently calculates the time-dependent accretion rate through the unstable zone. We further assume that this accretion rate directly feeds the inner ADAF, namely $\dot{m}_{\rm ADAF}=\dot{m}_{\rm u}$. To explore whether this scenario can reproduce the observed variability of SDSS J1430, {we consider a set of model parameters with $M_{\rm BH}=5\times10^{7}\rm M_\odot$, $\dot{m}_{\rm disk}=0.05$, $R_{\rm u}=15R_{\rm g}$, and $\alpha=0.25$. We find that, with these parameters, prescribing an appropriate time evolution of the unstable-zone width can reproduce the general trend of the observed optical variability.} Specifically, we assume that $\Delta R_{\rm u}$ remains at $0.05R_{\rm u}$ (i.e. $0.75\,R_{\rm g}$) during the first $\sim400$ days and subsequently decreases exponentially to $0.004R_{\rm u}$ (i.e. $0.06\,R_{\rm g}$) over the following $\sim520$ days. With this prescribed evolution, the model produces four prominent flares in $\dot{m}_{\rm ADAF}$, with the flare duration progressively shortening from $\sim400$ to $\sim$100 days, while their amplitudes also decrease, broadly resembling the observed behavior of SDSS J1430 (Figure \ref{fig:rim}C).

The resulting variable accretion rate then modulates the luminosity of the inner ADAF, which in turn irradiates the outer cold accretion disk and drives correlated optical variability \citep{George1991MNRAS.249..352G,Cackett2021iSci...24j2557C}. Since the optical emission region extends over hundreds of $R_{\rm g}$, we approximate the ADAF as a point source located at a height of $H=5R_{\rm g}$ above the disk \citep{Shakura1973A&A....24..337S}. Under these assumptions, we adopt a disk albedo of 0.5 and calculate the time-dependent optical emission from the irradiated disk \citep{George1991MNRAS.249..352G,Loska2004MNRAS.355.1080L,Cackett2007MNRAS.380..669C}, which is detailed in Appendix \ref{app:irr}. Adding the host galaxy contribution derived from the spectral decomposition in Section \ref{subsec:spectra type}, we obtain the predicted optical light curve in the observed frame, as shown in Figure \ref{fig:rim}B. The resulting light curve broadly reproduces both the observed trend of decreasing flare duration and amplitude. 

Nevertheless, the simplified toy RPI model has several notable limitations in reproducing the observed behavior of SDSS J1430. {The predicted flare magnitudes are not fully consistent with the observations, particularly for the first two flares, where the peak fluxes are poorly reproduced.} Furthermore, the predicted flux variations for the first two flares are predominantly concentrated near the peaks, whereas the evolution away from the peaks is minimal. In terms of photometric amplitude, the predicted $r$-band flux remains $\sim0.8$ mag fainter than the observed flux even though the host galaxy contribution is included, whereas the predicted $g$-band flux is roughly in agreement with the observation except for the first flare. 
Moreover, the progressive shrinking of the unstable zone assumed to reproduce the observed shortening of the flare durations is prescribed rather than independently predicted by the model. The underlying physical mechanism driving this contraction remains unclear. Current numerical models also do not yet incorporate a time-dependent warm corona or blurred reflection, which introduces an additional caveat in interpreting the evolving flare properties within this framework. The model also does not account for the long-term secular decline in accretion rate (and hence luminosity) observed in the optical light curve. {Therefore, while this calculation serves as a proof of concept showing that radiation pressure instability can qualitatively reproduce some of the characteristic variability of SDSS J1430, the limitations discussed above prevent us from establishing it as a definitive explanation.}

We also note that the slight optical brightening between approximately 2014 and 2018 suggests a moderate rise in the accretion rate, which likely triggered the radiation pressure instability. However, we cannot address why the accretion rate increased during this period by roughly a factor of two relative to the flux level in the low state. This increase may also alter the accretion states and drive CL AGNs to evolve from low state to high state \citep{Ai2020ApJ...890L..29A,Jin2021ApJ...912...20J,Guolo2021MNRAS.508..144G}. We also note that, after MJD $\sim60500$, the optical flux shows a slight rebrightening, followed by a strong flare with an amplitude comparable to that of the first flare (MJD 58500–58900). This may indicate the onset of another cycle of the radiation pressure instability.


\section{Conclusions}
\label{sec:conclusions}

In this paper, we present a comprehensive multi-wavelength study of the CL AGN SDSS J1430, combining optical, UV, and X-ray observations spanning from 2005 to 2026. Optical spectroscopic observations reveal a transition from Seyfert 1.9 to Seyfert 1.2, accompanied by the emergence of broad H$\beta$ emission. During the high state, the optical light curve exhibits at least four prominent flares, with their durations decreasing from $\sim400$ to $\sim$100 days over three years, accompanied by a progressive damping in the flare amplitude. Our main conclusions are as follows: 

\begin{enumerate}

    \item The long-term color variation follows the ``bluer-when-brighter'' trend, with a color–magnitude slope consistent with previous statistical results for CL and Type 1 AGNs.

    \item The UV and X-ray emission show significantly more dramatic variations with UV flare reaching amplitudes of $\sim0.5$ mag and X-ray flux varying by up to an order of magnitude. {In the most recent observations after MJD 60700}, a renewed optical flare with an amplitude comparable to the initial event is accompanied by X-ray and UV enhancements exceeding any previously recorded levels.

    \item {Both the warm corona and blurred reflection models provide comparably good fits to X-ray spectra. While the available observations do not allow us to unambiguously distinguish between the two scenarios, the weak Fe K$\alpha$ feature makes the warm corona more plausible.}

    \item {X-ray spectral fitting reveals a weak soft excess, with the soft excess strength $q$ generally below 0.04 for most observations. We find both $q$ and $\Gamma$ positively correlated with $\lambda_{\rm Edd}$, broadly following relations reported for Type 1 AGNs, although SDSS J1430 generally exhibits a weaker soft excess at a comparable Eddington ratio for most observations. }

    \item X-ray timing analysis reveals a low-frequency hard lag $\sim 10^{-4}$ Hz in both the brightest and the combined faint observations, with consistent amplitudes within $1\sigma$ uncertainties throughout the decline. The stability of this hard lag indicates that the disk-corona geometry remains essentially unchanged during the fading phase.

    \item {We explore radiation pressure instability as an illustrative possibility for the optical oscillations. A toy model with a prescribed shrinking unstable zone qualitatively reproduces the decreasing flare durations and amplitudes, although it does not fully capture the observed light curve morphology.}
\end{enumerate}

The multiple-flare behavior observed in SDSS J1430 is not unique, as similar long-term optical oscillations have been reported in other AGNs (e.g. \citealt{Ward2024ApJ...961..172W}). In future work, we will expand the sample of CL AGNs exhibiting such recurrent flaring behavior and systematically investigate the multi-wavelength properties of these sources, including the evolution and strength of their soft excesses and the possible role of radiation pressure instability in driving their long-term variability. Such a sample study will help establish whether the properties identified in SDSS J1430 are representative of a broader population of CL AGNs.

\begin{acknowledgements}

We thank Wei Yu for the helpful discussion on the timing analysis.
B.Y.\ is supported by ``the Fundamental Research Funds for the Central Universities''; by NSFC grants 12322307, 12273026, and 12361131579;
Xiaomi Foundation / Xiaomi Young Talents Program; The data analysis in this paper have been done on the supercomputing system in the Supercomputing Center of Wuhan University. MS acknowledges the Czech Science Foundation (GA\v{C}R) grant no. 26-23342I. BC has received funding from the European Research Council (ERC) under the European Union’s Horizon 2020 research and innovation program (grant agreement No. [951549]). BC and MS acknowledge the Czech-Polish Mobility program of the two Academies of Sciences, titled ``Appearance and dynamics of accretion onto black holes''. CR acknowledges support from SNSF Consolidator grant F01$-$13252 and ANID BASAL project FB210003. XC acknowledges support from the NSFC (12533005, 12233007, and 12347103), the science research grants from the China Manned Space Project with CMS-CSST-2025-A07.
\end{acknowledgements}


\bibliographystyle{aa}
\bibliography{references}

\clearpage

\begin{appendix}
\label{appendix}

\section{Estimating the emitting radius from the blackbody normalization}
\label{app:bbody}
{The normalization in model {\tt bbody} is defined by 
\begin{equation}
    N_{\rm bb}=\frac{L_{39}}{D_{10}^2},
\end{equation}
where $L_{39}$ is the source luminosity in units of $10^{39}\,\rm erg\,s^{-1}$ and $D_{10}$ is the distance of the source in units of 10 kpc \citep{Arnaud1996ASPC..101...17A}. Thus, the luminosity of the blackbody component can be estimated as
\begin{equation}
    L_{\rm bb}=N_{\rm bb}\times10^{39}D_{10}^{2}\ \ \rm{erg\,s^{-1}}.
\end{equation}
Assuming isotropic blackbody emission, the corresponding emitting radius is obtained from the Stefan–Boltzmann law,
\begin{equation}
    R_{\rm bb}=\left(\frac{L_{\rm bb}}{4\pi\sigma T^{4}}\right)^{1/2},
\end{equation}
where $\sigma$ is the Stefan–Boltzmann constant and $T$ is the blackbody temperature, converted from the fitted $kT$ (in keV) using $T=1.1605\times10^{7}(kT/\mathrm{keV})\ \mathrm{K}$. The radius is then expressed in units of the gravitational radius, $R_{\rm g}=GM_{\rm BH}/c^{2}$. For the fitted normalization of $N_{\rm bb}\sim5\times10^{-6}$ and $kT_{\rm bb}\sim72\, \rm eV$, the inferred emitting radius is approximately $R_{\rm bb}\sim0.01R_{\rm g}$, which is physically implausible for a blackbody-emitting region around the central black hole.}

\section{Effect of the optical depth on the spectral fitting results}
\label{app:diff_tau}
\begin{figure}
    \centering
    \includegraphics[width=1\linewidth]{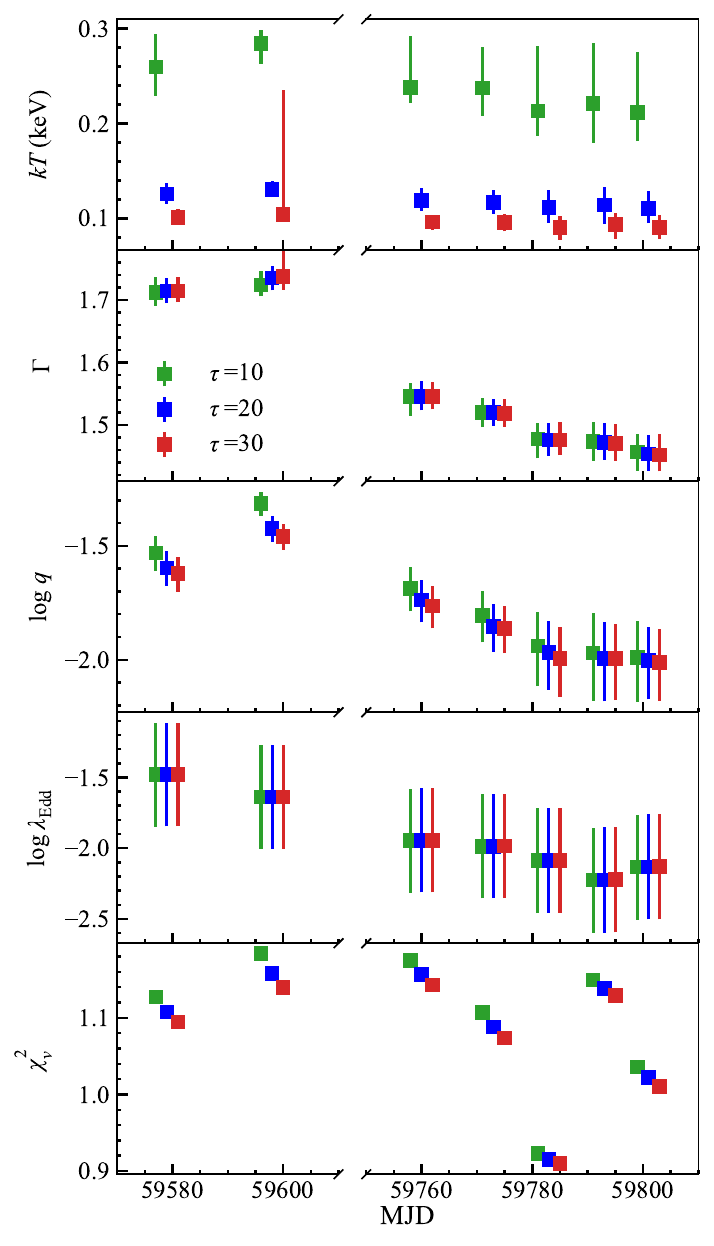}
    \caption{Temporal evolution of the spectral parameters obtained with the warm corona model {\tt tbabs*(comptt+nthcomp)} for different fixed optical depths. The red, green, and blue squares represent $\tau=10$, 20, and 30, respectively. For clarity, the data points for $\tau=10$ and 30 are shifted by $-2$ and $+2$ days, respectively.}
    \label{fig:diff_tau}
\end{figure}

{The optical depth $\tau$ of the warm corona is suggested to be in the range of $\tau\sim10-30$ \citep{Jin2012MNRAS.420.1825J}. To examine its effect on our results, we repeat the fits using the model {\tt tbabs*(comptt+nthcomp)}, fixing $\tau$ at 10 and 30, respectively. The resulting parameter evolution is shown in Figure \ref{fig:diff_tau}. We find that all models yield good fits for all spectra, and changing $\tau$ mainly introduces a systematic shift in the $kT_{\rm e}$, while $kT_{\rm e}$ remains consistent across epochs within the uncertainties for each $\tau$. In contrast, the evolution of the other key parameters, including $\Gamma$, $q$, and $\lambda_{\rm Edd}$, is largely insensitive to the value of $\tau$. Therefore, our main conclusions regarding the spectral evolution and accretion properties of the source are robust against the choice of $\tau$.}

\section{Unfolded spectra for all epochs}
\label{app:eeuf_all}
\begin{figure}
    \centering
    \includegraphics[width=1\linewidth]{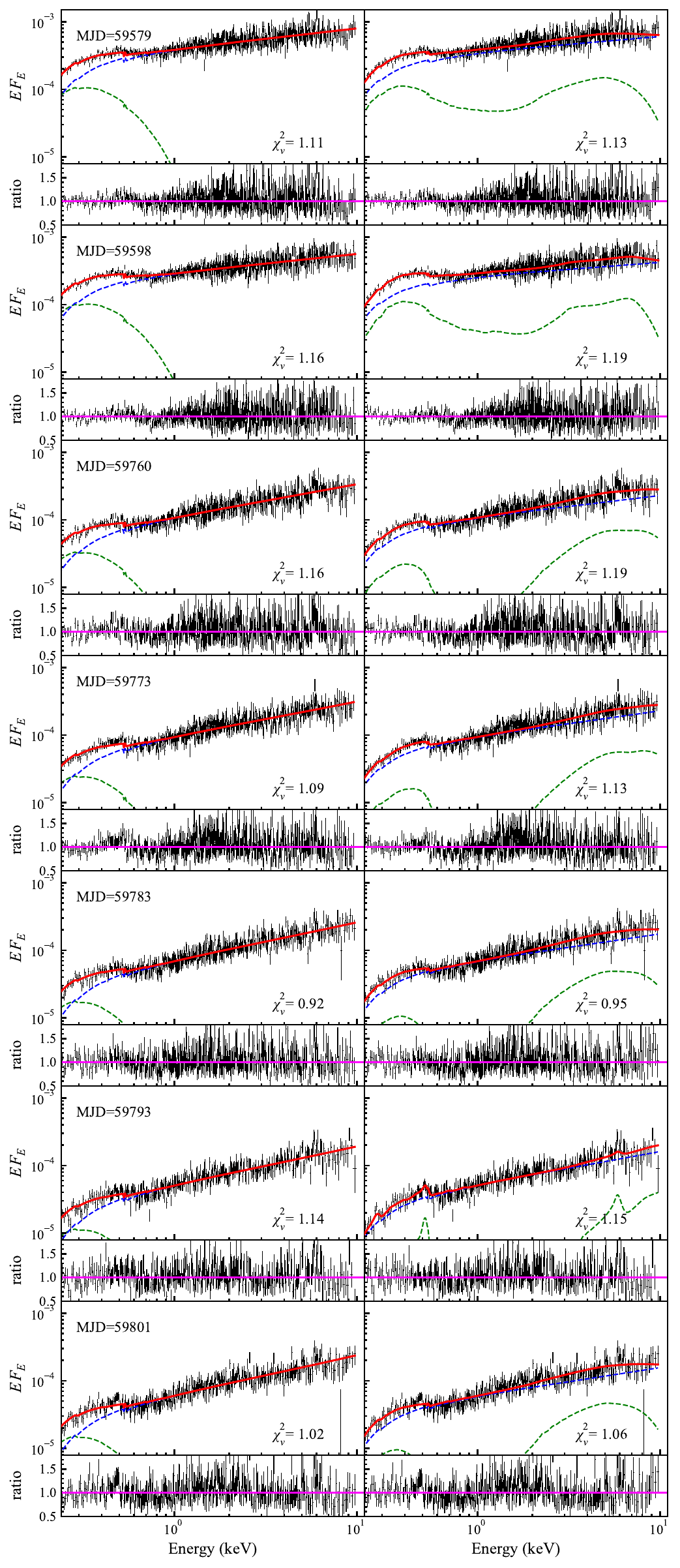}
    \caption{Unfolded spectra of SDSS J1430+2303 at all 7 epochs, fitted with the warm corona (left) and blurred reflection (right) models. The lower panels show the data-to-model ratios, with the magenta solid line marking ratio=1.}
    \label{fig:eeuf_all}
\end{figure}

{In this section, we present the unfolded XMM-PN spectra for all 7 epochs, fitted with the warm corona and blurred reflection models. The corresponding best-fitting models are shown in Figure \ref{fig:eeuf_all}.}

\section{Irradiation model for the reprocessed optical/UV emission}
\label{app:irr}
The high-energy photons emitted by the inner, luminous ADAF illuminate and heat the surface of the outer thin disk, which then re-radiates this reprocessed energy as optical continuum emission. This is the physical picture captured by the irradiation model. For a SMBH with $M_{\rm BH}=10^{7}\sim10^8\,\rm M_{\odot}$ and $\dot{m}=10^{-2}\sim10^{-1}$, the optical continuum is emitted at $10^2\sim10^3\,R_{\rm g}$ \citep{Shakura1973A&A....24..337S}. Therefore, we treat the inner ADAF as a point source in the lamppost geometry, located on the disk rotation axis at height $H = 5\,R_g$ above the disk plane. Given the tendency of SDSS J1430 to host a rapidly spinning BH, we adopt an inner disk radius $R_{\rm in}=2\,R_{\rm g}$ and outer radius $R_{\rm out}=10^5\,R_{\rm g}$.

The bolometric luminosity of the ADAF can be estimated using
\begin{equation}
    L_{\rm ADAF}=\eta\dot{M}_{\rm ADAF}c^2
\end{equation}
where
\begin{equation}
\begin{aligned}
& \dot{M}_{\rm ADAF} = \dot{m}_{\rm ADAF}\dot{M}_{\rm Edd}, 
\quad
\dot{M}_{\rm Edd} = \frac{L_{\rm Edd}}{\eta c^2},\\
& L_{\rm Edd} = 1.26\times10^{38} \left(\frac{M_{\rm BH}}{M_\odot}\right)\,{\rm erg\,s^{-1}}.
\end{aligned}
\end{equation}
Here $\eta$ is the radiation efficiency, $\dot{M}_{\rm Edd}$ is the Eddington accretion rate and $L_{\rm Edd}$ is the Eddington luminosity. Under the point-source approximation, and assuming the ADAF emits isotropically, the irradiating flux received by the disk at $R$ is then
\begin{equation}
F_{\rm irr}(R) = \left(1-A\right)\frac{L_{\rm ADAF}}{4\pi}\,
\frac{H}{\left(R^2+H^2\right)^{3/2}},
\end{equation}
where $A$ is the disk albedo \citep{George1991MNRAS.249..352G,Loska2004MNRAS.355.1080L,Cackett2007MNRAS.380..669C}. The local viscous dissipation flux of the thin disk is defined by \citep{Shakura1973A&A....24..337S}
\begin{equation}
F_{\rm visc}(R) = \frac{3 G M_{\rm BH} \dot{M}_{\rm disk}}{8\pi R^3}
\left[ 1-\sqrt{\frac{R_{\rm in}}{R}} \right].
\end{equation}
Hence the effective temperature of the disk surface at $R$ under irradiation follows,
\begin{equation}
T_{\rm eff}(R) = \left[\frac{F_{\rm visc}(R)+F_{\rm irr}(R)}{\sigma}\right]^{1/4},
\end{equation}
where $\sigma$ is the Stefan--Boltzmann constant. Then, the monochromatic luminosity in the rest-frame can be computed by integrating the local blackbody emission, weighted by $T_{\rm eff}(R)$ over the disk surface, which is 
\begin{equation}
L_\nu^{\rm rest}(\nu) = 4\pi^2 \int_{R_{\rm in}}^{R_{\rm out}}
R\,B_\nu\!\left[\nu, T_{\rm eff}(R)\right]\,{\rm d} R .
\end{equation}
where 
\begin{equation}
    B_\nu=\frac{2h\nu^3/c^2}{e^{h\nu/k_{\rm B}T}-1}
\end{equation}
is the Planck function, $h$ is the Planck constant and $k_{\rm B}$ is the Boltzmann constant. Given the filter transmission ranging between $\nu_{\rm obs,\,min}$ and $\nu_{\rm obs,\,max}$ in the observer frame, the observed flux density at the effective frequency of the filter, $\nu_{\rm obs,\,eff}=c/\lambda_{\rm obs,\,eff}$, is obtained by applying the appropriate redshift (K-)correction to $L_\nu^{\rm rest}$,
\begin{equation}
f_\nu^{\rm obs}(\nu_{\rm obs,\,eff}) \approx \frac{(1+z)\,L_\nu^{\rm rest}\!\left[(1+z)\,\nu_{\rm obs,\,eff}\right]}{4\pi d_L^2},
\end{equation}
with $d_L$ the luminosity distance and $z$ the redshift. Rather than convolving $L_\nu^{\rm rest}$ with the full filter transmission curve, we evaluate $f_\nu^{\rm obs}(\nu_{\rm obs,eff})$ at this single effective frequency for computational simplicity. The predicted AB magnitude then follows as
\begin{equation}
m_{\rm AB} = -2.5\,\log_{10}\!\left(\frac{f_\nu^{\rm obs}}{3631\ {\rm Jy}}\right).
\end{equation}

For SDSS J1430, we assume $\eta=0.1$, $A=0.6$, $M_{\rm BH}=5\times10^{7}\,M_\odot$ and $\dot{M}_{\rm disk}=0.05\,\dot{M}_{\rm Edd}$. $\dot{M}_{\rm ADAF}$ is taken to be equal to the time-dependent output accretion rate of the unstable zone derived from the RPI model. Given the frequency ranges of the ZTF $g$ and $r$ filters \footnote{see https://svo2.cab.inta-csic.es/theory/fps/index.php}, the rest-frame disk region dominating the emission in each band can be approximated using the Wien displacement relation,
\begin{equation}
\frac{h\nu_{\rm rest}}{k_{\rm B}T_{\rm eff}(R)} = b_\nu,
\qquad
\nu_{\rm rest}=(1+z)\nu_{\rm obs}
\end{equation}
where $b_\nu=2.82144$ is the Wien displacement constant in frequency form. Once the dominant emission region is identified for each band, the predicted magnitude variation can be computed using the effective wavelength (or effective frequency) of that band.

\end{appendix}

\end{document}